\documentclass[preprint,12pt]{elsarticle}

\usepackage{amssymb}
\usepackage{amsmath}
\usepackage{graphicx}
\usepackage{epsfig}
\usepackage{color}
\definecolor{red}{rgb}{0.00,0.00,0.00}

\journal{Progress in Surface Science}
\begin{document}


\begin{frontmatter}

\title{Excitation of local magnetic moments by tunnelling electrons}
\author[1]{Jean-Pierre Gauyacq}
\address[1]{
Institut des Sciences Mol\'eculaires d'Orsay, ISMO, Unit\'e mixte CNRS-Universit\'e Paris-Sud,UMR  8214, B\^atiment 351, Universit\'e Paris-Sud, 91405 Orsay CEDEX,
France}
\author[cin2]{Nicol\'as Lorente}
\author[cin2]{Frederico Dutilh Novaes}
\address[cin2]{Centro de investigaci\'on en nanociencia y nanotecnolog\'{\i}a
 (CSIC - ICN), Campus de la UAB,
     E-08193 Bellaterra, Spain}
\date{\today}
\begin{abstract}

The advent of milli-kelvin scanning tunneling microscopes (STM) with
inbuilt magnetic fields has opened access to the study of magnetic
phenomena with atomic resolution at surfaces. In the case of single atoms adsorbed on a surface,  
the existence of different magnetic energy levels localized on the adsorbate is due to the  breaking of  the rotational invariance of the adsorbate spin by the interaction with its environment, leading to  energy terms in the meV range.
These structures were revealed by
 STM experiments in IBM Almaden in the early 2000's 
for atomic adsorbates on CuN surfaces. The experiments consisted in the study of the changes
in conductance caused by inelastic tunnelling of electrons (IETS, Inelastic Electron Tunnelling Spectroscopy). Manganese
and Iron adatoms were shown to have
different magnetic  anisotropies induced
by the substrate. More experiments by other groups followed up, showing that magnetic
excitations could be detected in a variety of systems: e.g. complex organic
molecules showed that their magnetic anistropy was dependent on
the molecular environment, piles of magnetic molecules
showed that they interact via intermolecular exchange interaction,
spin waves were excited on ferromagnetic surfaces
and in Mn chains, and  magnetic
impurities have been analyzed on  semiconductors. 
These experiments brought up some intriguing questions: the efficiency of
magnetic excitations was very high, the excitations  could or could not involve 
spin flip of the exciting electron  and singular-like behavior was
sometimes found at the excitation thresholds. These facts called for extended
theoretical analysis;  perturbation theories,  sudden-approximation approaches and a strong coupling  scheme successfully explained
most of the magnetic inelastic processes. In addition, many-body approaches were also used to decipher the interplay between inelastic processes and the Kondo effect.  
Spin torque transfer has been shown to be effective in changing spin orientations of an adsorbate in theoretical works,
 and soon after it was shown 
experimentally.
More recently, the previoulsy mentioned strong coupling approach was extended to treat the excitation of spin waves in atomic chains and the 
ubiquitous role of electron-hole pair creation in de-exciting spins on surfaces
has been analyzed. 
This review article expounds these works, presenting
the theoretical approach by the authors while trying to thoroughly
review parallel theoretical and experimental works. 

\end{abstract} 
\begin{keyword}
IETS \sep magnetism \sep spin flip \sep MAE \sep magnetic anisotropy\sep spin-orbit coupling \sep scanning tunneling microscope \sep STM
\sep inelastic effects \sep conductance \sep electron transport 
\sep spectroscopy \sep magnetic adsorbates \sep lifetimes \sep Kondo effect
\end{keyword} 
\end{frontmatter}

\bibliographystyle{elsarticle-num}

\section{Introduction}

Tunneling phenomena is a purely quantal phenomena with
great impact in current basic and applied research.
The advent of solid-state devices led to the study
of electron tunneling through insulating barriers in order to create
a pletora of device designs based on electron tunneling~\cite{Esaki}.
From the fundamental point of view, tunneling offered many interesting
phenomena and applications from imaging of surface structure and
topography~\cite{STM} to Josephson effect~\cite{Josephson} and to
inelastic electron tunneling spectroscopy (IETS)~\cite{Jacklevic}.

Inelastic electron tunneling spectroscopy was discovered when studying
electron tunneling through an insulating thin film between two metallic
electrodes. Jacklevic and Lambe ~\cite{Jacklevic} recorded differential conductance traces
where a rich structure appeared at certain well-defined voltages. Their
analysis led to the conclusion that they were measuring the change
in conductance due to the excitation of vibrations of unknown
impurities in the insulating layer. This finding led to the creation
of a new type of spectroscopy, IETS, that was \textcolor{red}{much} 
 developed in the
70's and 80's. Hansma~\cite{Hansma} summarizes in a very interesting
review article many of the molecular species studied in this
way in different types of tunneling barriers, and on-going
research efforts are currently undertaken in the IETS of
insulating layer interfaces~\cite{Petit}.

The advent of the scanning tunneling microscope (STM) started the search
of IETS in the tunneling junction of the STM~\cite{Garcia}.
The stakes were high: on the one hand-side, STM would be able
to detect the vibrational signatures of the species in the junction
making it possible to develop a chemical sensitivity absent in the usual
STM operational modes; on the other hand-side, the extreme local sensitivity
of the STM would permit to have a single-molecule spectroscopy. Despite
theoretical evaluations that IETS was within reach in STM~\cite{Garcia,PerssonPRL1987},
experimental proof only came in 1998 when Stipe, Rezaei and Ho
showed the vibrational IETS of a single acetylene molecule adsorbed
on a Cu (100) surface~\cite{StipeScience1998}. There are excellent review articles
that
describe the physics and history of IETS with the STM~\cite{Hiromu,Komeda}.

The sophistication of STM opens the possibility of addressing lower
energy scales. Very low temperatures and extreme sensitivity
equipment appear along the 90's. Once that vibrational IETS was
proven with the STM, lower-energy excitations became available. In 2004,
Heinrich and co-workers~\cite{HeinrichScience2004} showed that magnetic
excitations on a single magnetic atom were detected using a milli-Kelvin
STM with a built-in magnetic field. This seminal experiment has given
rise to a lot of activity in magnetic IETS
on the atomic scale.

\begin{figure}
\centering
\includegraphics[width=0.8\textwidth]{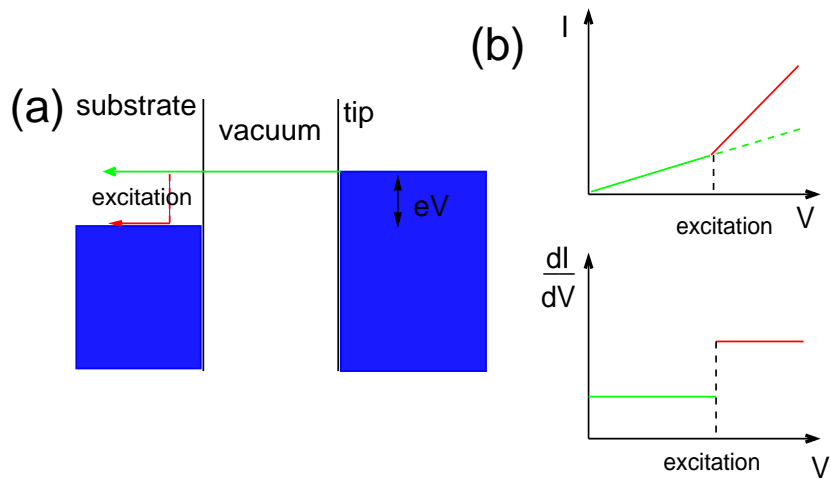}
\caption{\label{schema_inelastique}
One-electron picture of the increase in conductance
when an excitation takes place in the tunneling
current. $(a)$ Level scheme depicting the substrate and STM's tip
electrodes as two metals with electron states
filled up to the Fermi energy, which are respectively
shifted by bias $V$ times the electron charge $e$. Electrons
can tunnel elastically (green trajectory) but also inelastically
when the bias is larger than the excitation energy, because
the inelastic trajectory proceeds above the substrate's
Fermi energy (red trajectory). $(b)$ The I--V characteristic
is roughly linear with the bias. When the bias matches the
excitation energy the inelastic channel becomes available. Hence
the current increases. The lower pannel shows the
differential conductance behavior which is flat but
presenting a discontinuity at the bias matching the excitation energy.
}
\end{figure}

Both vibrational and magnetic IETS consist in a measurable change
of conductance due to an excitation of an atom, molecule or general
atomic structure under the tip of an STM. Hence, the tunneling
current is both the exciting and the measuring probe. This dual
behavior of the tunneling current makes IETS a complex technique
where a simple-minded picture is surely error prone. However, a
first-order approximation of how the excitation of an atom or molecule
changes the conductance can be easily found in terms of the opening
of new conduction channels linked to the excited molecular 
states~\cite{Jacklevic,Hansma}, see Fig.~\ref{schema_inelastique}. Indeed, when the tip-sample bias
is larger than the excitation energy, the tunneling electron can
cede part of its energy to the molecule and still end up in a state
above the Fermi energy of the corresponding electrode, thus contributing to the tunnelling current as part of an inelastic current. For
bias below this threshold, the
tunneling current is just formed of elastic electrons. When the new
channel opens at the bias matching the excitation energy, the tunneling
current increases because it now contains elastic as well as inelastic
electrons. This abrupt change in the current leads to a jump
in the differential conductance, and to a peak in the second
derivative of the current with respect to bias, centered about the
excitation energy in eV. 

As announced, the previous picture is simplistic and it does not take
into account the complexities of the many-body character of the excitation
process. There are cases where the conductance decreases instead of
increasing. This was shown by Hahn {\em et al.}~\cite{HahnPRL2000}
in the case of O$_2$ adsorbed on Ag (110) where the IETS
shows dips instead of peaks. The study of this system showed that
the appearance of dips can be associated with the mixed-valence
electronic structure character of O$_2$ on Ag (110)~\cite{AlducinPRL2010}.

A big difference between vibrational and magnetic IETS was quickly revealed.
While vibrational IETS rarely implies increases of the conductance of
more than 10\%, magnetic IETS easily reaches 
100\% or more of change in conductance. This behavior can be traced back to the
strengths of the interactions at play: while electron-vibration couplings
are weak, electron-spin couplings are very large. The first consequence
of this fact is that the perturbational approaches developed for
vibrational IETS~\cite{PerssonPRL1987,MingoPRL2000,Ueba,LorentePRL2000,LorenteAPA2004} are no longer valid. A second consequence is \textcolor{red}{that multiple successive excitations
are easily accessible in magnetic IETS~\cite{LothNatPhys2010} in the case of strong currents}. Finally,
vibrational IETS is very sensitive to the symmetry and to the
particular system and thus, only a few modes are detectable which
has led to the creation of propensity rules for mode analysis~\cite{Ratner,PaulssonPRL2008}. 
Similarly, not all excited states in a magnetic system can be excited by a tunnelling electrons (see e.g. below the discussion on spin wave excitation in Heisenberg chains) 
and this can be easily rationalized
in terms of angular momentum conservation  and spin-coupling coefficients~\cite{Lorente_PRL2009}.

Magnetic IETS has not  been confined to a small number of atomic
systems, but it has also been extended to spinwave excitation~\cite{BalashovPRL2006} and
itinerant magnetism~\cite{KhajetooriansPRL2011}. Moreover, 
the use of spin-polarized STM has made a natural
connection of magnetic excitations with
spin torque of magnetic atomic systems. The 
 reversal of the magnetization
of atomic structures by the tunneling current
has been shown on magnetic islands~\cite{KrauseScience2007,HerzogAPL2010}
and  also on single atoms both at the 
experimental~\cite{LothNatPhys2010} and theoretical~\cite{DelgadoPRL2010,NovaesPRB2010} levels.

Finally, magnetic IETS is intrinsically linked to the Kondo effect. Indeed, Kondo effect is induced by spin-flip transitions in an impurity induced by collisions with the substrate electrons; this is exactly the same process as the one at play in magnetic IETS and one can expect strong links between the two phenomena, as well as the possible emergence of many-body effects (Kondo-like effects) in magnetic IETS.

Hence, despite
its short life, magnetic IETS is a well established technique
that we will expound in the present article.
First, we will review the main experimental results that
have been briefly mentioned above, together with other results 
to give the reader a vision of the breadth of the field. Second, we
will review and explain the main theoretical approaches trying to
emphasize the main features of magnetic IETS. Finally, we will conclude
and try to outline some perspectives of this powerful technique.

%
%
%
%
%
%

\section{Experimental Results}
\label{experiments}

We will review the experimental work on magnetic IETS by considering particular
systems and the physics explored in those systems rather than by proceeding in chronological order. We aim at presenting the experiments in such a way as
their particular features are emphasized.

In order to achieve this we will first review the experiments performed on
atomic adsorbates decoupled from the metallic substrate by an atom-thick
insulating layer. The presence of a thin insulating layer between adsorbate and substrate effectively separates the two and 
leads to their partial decoupling, and thus to longer electronic lifetimes for states localised on the adsorbate. Experimental studies of these excited states become easier and better resolved and even 
spin lifetimes become measurable as we will present. \textcolor{red}{Yet, IETS succeded in revealing the presence of magnetic excitations for magnetic atoms adsorbed directly 
on a metallic substrate. }
Metallic substrate can also be magnetic, given rise to collective
excitation such as magnons. We will also review these experiments.
Other type of collective excitations can take place in chains of adsorbates
as has been shown for artificially assembled Mn chains. Next, we will increase
the complexity of IETS by studying the experiments performed with
polarized electrons that naturally lead to spin torque and magnetization
reversal. Finally, we will briefly mention experiments on magnetic IETS
and Kondo physics.

\subsection{Magnetic adsorbates partially decoupled from the metallic substrate}
\label{decoupled}

Heinrich and co-workers~\cite{HeinrichScience2004} showed for the first
time that the STM could be used to create a magnetic excitation in an
atomic adsorbate. The experiment was performed in ultra-high vacuum (UHV),
at 0.6 K, and in the presence of a magnetic field. The
system was an adsorbed manganese atom on two layers  of alumina (Al$_2$O$_3$)
on a crystalline substrate of nickel aluminum. With this setup, they
measured the bias at which an inelastic channel opens and the tunneling
conductance presents a well-determined step. They used a high magnetic
field to orientate the manganese spin. The measured energy needed to flip the
adatom spin can be evaluated as the Zeeman energy of a simple local magnetic moment:
\begin{equation}
\Delta = g \mu_B B
\label{Zeeman}
\end{equation}
where $\mu_B$ is the Borh magneton, $B$ the magnetic field and
$g$ the gyromagnetic factor. $\Delta$ is a very small energy. For
$B=7$ T, $\Delta \approx 0.8$ meV. This can only be resolved
if the temperature  is   in the milli-Kelvin range (presently 600 mK) 
and if the adatom is
substantially decoupled from the electron-hole pair excitations of the
substrates in order to have a well-defined excitation energy. This extended
lifetime is achieved by decoupling the adatom from the substrate by
an insulating film of Al$_2$O$_3$.

The experiments measured $\Delta$ as a function of the applied magnetic
field by measuring the bias at which a step was found in the STM conductance.
From Eq.~(\ref{Zeeman}), the gyromagnetic factor $g$ was found to vary
between 1.88 and 2.01 depending on the location of the Mn atom
on the Al$_2$O$_3$ island. This first magnetic IETS example showed that
$g$ depends on the local environment of the studied atom, hinting at the existence of local interactions 
influencing the adsorbed magnetic atom.

In the absence of a magnetic field, the magnetic excitations of a free Mn
atom are in the range of eV, corresponding to typical values
of the exchange energy in atoms. However, on a substrate, the measured energies
of excitations in adsorbed magnetic atoms correspond well
to the Zeeman splitting plus the surface imposed magnetic anisotropy. 
Hence, typical adsorbate magnetic excitations are in the range of meV rather
than eV.

\begin{figure}
\centering
\includegraphics[width=0.8\textwidth]{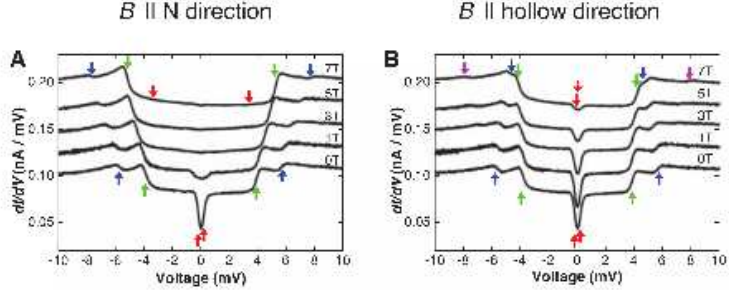}
\caption{\label{Fe}
Conductance measurement for an STM junction on a Fe adatom on CuN/Cu(100),
Ref.~\cite{HeinrichScience2007}. The Fe adatom is adsorbed ontop
a Cu atom. 
\textcolor{red}{A magnetic B field is applied along two different directions parallel to the surface: a direction in the plane  containing a Fe adsorbate and  one of its first N neighbours ('N-direction') and the perpendicular direction ('hollow direction').}
{As a consequence of the anisotropic distribution of N neighbours around the Fe adsorbate, the  magnetic structure of the Fe adsorbate  shows a very clear anisotropy. It is
revealed in the conductance spectrum when the external magnetic
field is aligned along the $A$ direction without or
$B$ with N atoms. The steps in the conductance take place
at different energies stressing the magnetic anisotropy.  From reference~\cite{HeinrichScience2007}}. Reprinted with permission from AAAS.
}
\end{figure}

Further experiments by the same group~\cite{HeinrichScience2007}
explored more systems, revealing MAE \textcolor{red}{(Magnetic Anisotropy Energy)} by magnetic IETS. Indeed, Mn on a
monolayer of CuN on Cu(100), and Fe on CuN on Cu (100), show excitation
energies in the meV range, which evidences the sizable MAE of these
systems. Figure~\ref{Fe} is the conductance measurement for the Fe/CuN/Cu
(100) as a function of applied bias for different magnetic fields. Let
us briefly present how MAE determines the low-energy spectra revealed
by magnetic IETS.

In iron-group atoms, the crystal-field splitting of the 2L+1 levels
(where L is the orbital quantum number) is much larger than the
spin-orbit coupling, even for the above case of adsorbates on a CuN
monolayer. This leads to the quenching of the orbital angular momentum,
see for example~\cite{Yosida}, and to the use of the spin operator to
determine the magnetic state of the adsorbate. 
The spin-orbit interaction couples the states split by the crystal field \textcolor{red}{to higher lying states} leading to second-order energy shifts of the states; it thus brings 
a coupling between the spin direction and its environment. The spin-orbit induced energy shifts can be represented by 
 the effective spin
Hamiltonian~\cite{Yosida}: 
\begin{equation} \hat{H}= D \hat{S}_z^2 + E
( \hat{S}_x^2 - \hat{S}_y^2 ).  
\label{anisotropy} 
\end{equation} 
The constants $D$ and $E$ depend on the spin-orbit coupling constant and
the crystal-field split energy levels of the adatom. Hence, MAE is due
to the effect of the environment on the adatom spin via the spin-orbit
coupling. The diagonalization of Hamiltonian~(\ref{anisotropy}) leads to
the actual spin states of the adsorbate and to the low-energy magnetic
excitation energies.

Hirjibehedin and collaborators~\cite{HeinrichScience2007} measured
the excitation energies of the above Fe/CuN/Cu(100) and Mn/CuN/Cu(100)
systems. From these energies they extracted $D$ and $E$, becoming the
first measurement of MAE and of the anisotropy Hamiltonian~(\ref{anisotropy})
at the single-atom level.

\textcolor{red}{They found } $D=-1.55$ meV and $E=0.31$ meV together
with $g=2.11$ for Fe and $D=-0.039$, $E=0.007$ and $g=1.90$ for adsorbed Mn. The
eigenstates of Hamiltonian~\ref{anisotropy} can now be found
and fully characterized the IETS. These eigenstates are given in
Ref.~\cite{HeinrichScience2007} in terms of the free-atom spin. It is
very instructive to look at the composition of the anisotropy states,
Table~\ref{tableFe}.  The excitation from the ground
state to the second excited state of Fe basically involves transitions
from $S_z=\pm 2$ to $S_z=\pm 1$
 hence
we can say that the impinging electron has to flip its spin to produce
the excitation since Fe spin changes in $|\Delta S_z|= 1$. However, 
excitations to the first excited level remains among the $|S_z|=2$
components 
Hence, this transition does
not involve a spin flip of the impinging electron. The mixture
of spin states by the anisotropies modifies any selection rule
based on the spin of the colliding electron~\cite{Lorente_PRL2009}. We
will further elaborate on this point in the theory section.

\begin{table}[hb]
  \begin{tabular}{|l|ccccc|}
\hline 
      & $|2,+2\rangle$& $|2,+1\rangle$& $|2,0\rangle$& $|2,-1\rangle$& $|2,-2\rangle$  \\ \hline
Ground state & 0.697 & 0 & -0.166 & 0 & 0.697 \\ \hline
First excited & 0.707 & 0 & 0 & 0 & -0.707 \\ \hline
Second excited & 0 & 0.707 & 0 & -0.707 & 0 \\ \hline
Third excited & 0 & 0.707 & 0 &  0.707 & 0 \\ \hline
Fourth excited & 0.117 & 0 & 0.986 & 0 & 0.117 \\ \hline
  \end{tabular}
  \caption{Coefficients of the spin states obtained after diagonalizing Hamiltonian Eq.~(\ref{anisotropy}) for a zero external magnetic field in the $S=2$
manifold.  The anisotropy due to the crystal field imposed
by the surface mixes up the different $S_z$ components. If an external magnetic
field is turned on, a priviledged direction is set, and the spin states
increasingly resemble the free atom states under the Zeeman effect for
larger magnetic fields. Taken from reference~\cite{HeinrichScience2007}.} 
  \label{tableFe}
\end{table}

\begin{figure}
\centering
\includegraphics[width=0.8\textwidth]{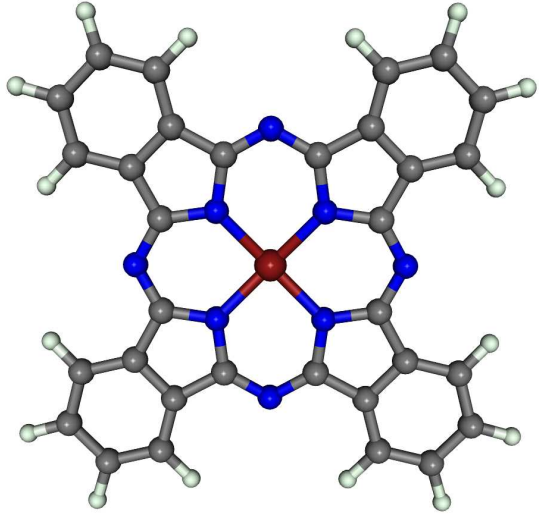}
\caption{\label{FePc}
 Ball-and-stick scheme of an iron phthalocyanine molecule.
 The central atom is the iron one, and the rest of atoms are nitrogen,
 carbon and hydrogen. 
 \textcolor{red}{The  free molecule has a $D_{4h}$ symmetry
which is reduced upon 
adsorption.} In the case of adsorption on
the CuO/Cu(110) substrate~\cite{TsukaharaPRL2009}, two different adsorption
geometries are compatible, leading to different symmetries and to
different MAE.
}
\end{figure}

Similar experiments have also been performed on magnetic molecules.
Tsukahara and coworkers~\cite{TsukaharaPRL2009} have measured magnetic
IETS on a iron phathalocyanine molecule adsorbed on a Cu (110)
with a single layer of oxide.
Again, the oxide layer reduces the charge transfer from the metal substrate
to the molecule and the molecule keeps much of its free molecule character.
However, adsorption has the dramatic consequence of reversing the 
anistropy sign. Indeed $D=8.9$ meV in the gas phase~\cite{Dale_1968a,Dale_1968b} and the study
of the magnetic IETS leads the authors to a $D=-6.4$ meV value.
The transversal anistropy $E$ is \textcolor{red}{very small.} 
Iron phathalocyanine is a $S=1$ molecule. The ligand field splits
the $d$-electron manifold of the Fe atom at the center of the molecule,
Fig.~\ref{FePc}, and the 6 electrons in the d-orbitals are now
rearranged following the tetrahedral energy ordering
of the split d-orbitals. As a consequence, two unpaired electrons lead
to a total $S=1$ in the molecule. 

The effect of the substrate is very important in the final valules
of the MAE. The authors find two possible orientations of the
molecule on the surface, leading to slightly different magnetic IETS
and to different MAE values, this is
clearly revealed under the effect of an external magnetic field. 
At zero magnetic field, the $S=1$ electronic structure leads to a magnetic
IETS showing only one step. This is because the strong $D$ and zero
$E$ values split the $S=1$ states into two levels of $|S_z|=1$ and
$S_z=0$. Hence, there is only one possible excitation. The experimental
inelastic change in conductance is 1/3 of the total conductance.  When the
magnetic field is ramped up, the  $|S_z|=1$ levels split, and one more
step appears in the magnetic IETS. Each of the two inelastic steps amount
to 25\% the elastic conductance. These values are characteristic of a
$S=1$ system~\cite{Novaes_FePc,K}.

\subsection{Lifetime measurements of adsorbate magnetic states}
\label{lifetime}

The existence of nano-magnets on a solid surface, the orientation of which
could be changed at will by tunneling electrons, opens fascinating
perspectives for the miniaturization of electronics. However, to lead
to easily manageable devices, the excitation of local spins must have,
among other properties, a sufficiently long lifetime. It is thus of
paramount importance to know the decay rate of the excited levels of
the local spin and in particular to decipher the various parameters
and effects that govern its magnitude. 

The magnetic IETS experiments described above
were performed in systems in which an insulating coating on the surface was
separating the magnetic adsorbate carrying the local spin from the metal
substrate. We will shortly review 
experiments on adsorbates directly deposited on a metallic
substrate that do not lead to sharp IETS structures~\cite{BalashovPRL2009} 
as the above ones
and this was attributed to a too short lifetime of the magnetic
excitation on metals, stressing the importance of the decoupling
layer between local spin and substrate in stabilizing the magnetic
excitation. 

De-excitation of a local spin implies energy transfer
from the local spin to the substrate degrees of freedom, i.e. to the
substrate electrons or phonons. Phonons are not directly coupled to
spin variables, but only via spin-orbit couplings which make
phonons particularly inefficient in the deexcitation process 
(see e.g. a discussion in \cite{FabianSarma1999}). In contrast, the
adsorbate spin variables can be directly coupled to substrate electrons
and electrons colliding on a magnetic adsorbate can easily induce
magnetic transitions. Actually, this is exactly what happens in the
magnetic excitation induced by tunneling electrons in the IETS experiments
described above; in the de-excitation process the tunneling electrons are
simply replaced by substrate electrons. The decay of excited magnetic
states in individual adsorbates thus proceeds via electron-hole pair
creation. 

Recently, the decay rate of excited magnetic Mn atoms adsorbed on
CuN/Cu(100) has been measured by Loth and collaborators~\cite{LothNatPhys2010}
via the analysis of the dependence of the adsorbate conductivity
on the tunneling current. By using magnetic inelastic excitation with
a tunneling current, they changed the population of 
the different magnetic states. This can be actively achieved by varying
the electron current. If the average time between tunneling electrons
is shorter than the excited state lifetime, 
then the tunnelling electrons are probing partly excited adsorbates instead of probing only ground state adsorbates.

This multiple excitation has a dramatic effect on the tunneling conductance.
Indeed, we have seen that an opening of an inelastic channel leads to
a sharp change in the conductance. Once the channel is open, the conductance
remains constant at low currents because the conduction conditions are
unaltered. However at higher currents, the average time between tunneling 
electrons will start matching the excited state lifetime. As a result
there will be tunneling electrons that probe the excited state instead
of only the ground state. The conductance of these excited states
is different leading to a current dependent conductance. Moreover, as
the bias increases, the number of inelastic electrons increase. Typically,
\textcolor{red}{in the Mn/CuN system studied by Loth et al~\cite{LothNatPhys2010}}
this leads to a drop of the conductance with bias after the first inelastic
threshold is matched. The conductance vs bias curve presents a peak instead
of step~\cite{LothNatPhys2010}.

Loth and collaborators~\cite{LothNatPhys2010} modelled their experiment
by using Pauli master equations or rate equations that consider
the population evolution of the different magnetic states as a function
of time depending on the excitation and deexitation rates. From
here they obtain that the typical spin lifetime of Mn on CuN/Cu(100) is
in the \textcolor{red}{sub-nano-second range}. Moreover,  the deexcitation rate depends
on the energy of the excited state since the phase space for electron-hole
pair excitation increases with the amount of available energy in the
deexciting atom. Hence, they find that for example, at $B=7$ T the longest
lifetime is 0.25 ns for the first excited state, but it is 0.73 ns at $B=3$ T
meaning a smaller deexcitation rate as the excitation energy decreases.

\begin{figure}
\centering
\includegraphics[width=0.8\textwidth]{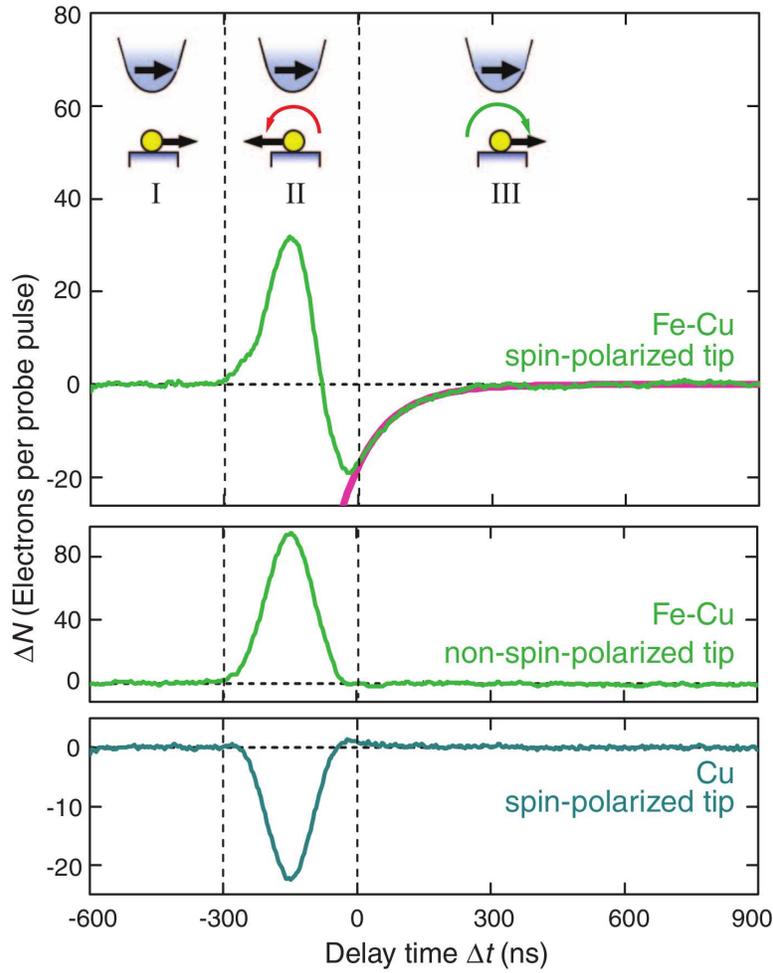}
\caption{\label{deltaN}
Pump-probe results with the STM. Here, 
the change in the number of detected electrons, $\Delta N$,
is measured as a function of the delay time. Experiments
are performed under an external magnetic field of 7 T. In the
region I, the
probe pulse precedes the pump pulse. In the region II, the pump and
probe pulses overlap, while in region III, the probe pulse follows
the pump pulse.  Insets depict the relative orientation of tip and
sample spins. For the Fe-Cu dimer (top panel) \textcolor{red}{$\Delta N$} decays exponentially
in region III, with a spin relaxation time of  87 ns obtained
from an exponential fit (magenta). Control experiments on the same
Fe-Cu dimer but without spin sensitivity in the tip (middle panel)
and on a Cu atom with spin-sensitivity in the tip (bottom panel). \textcolor{red}{From reference~\cite{LothScience2010}. Reprinted with permission from 
AAAS.}
 } 
\end{figure}

The same group recently presented a more direct way of measuring the
spin relaxation times~\cite{LothScience2010}. They used an all-electronic
pump-probe measurement scheme with an STM. A strong bias pulse first
excites the magnetic atom and a second weaker pulse probes
the excitation state of the same atom at a delayed time $\Delta t$. 
They studied the change in the number of detected electrons $\Delta N$
after the probe pulse compared to the number of tunneling
electrons for the atom in
its ground state. The number of electrons is obtained by integrating the
measured current over $\Delta t$. The number of electrons in the
ground state is obtained by integrating the current for a a probe
pulse that precedes the pump pulse for a given large time (600 ns). 
They used a spin-polarized STM tip with its axis aligned with the studied
atom. In the present case Fe on CuN/Cu(100). The excitation from the
first electron pulse changes the alignment of the atom spin with respect to the tip's,
implying a drop in the number of tunneling electrons. As the excited
state decays, the number of electrons increases. Hence, $\Delta N$ is
first negative and it goes exponentially to zero, since the current will
match the elastic current at large time, Fig.~\ref{deltaN}.
From this exponential decay, they obtained the excited state lifetime.

They increased the MAE of the adsorbed Fe atom by adding a nearby Cu atom.
In this way, the first excitation energy was increased by a factor 4 as
compared to the singly adsorbed Fe atom. As a consequence of the large
anisotropy, the measured spin lifetime was $\sim 50$ ns for
$B=1$ T, much larger than the relaxation times found for  single Mn
adsorbates. 

These experiments show that an insulating layer and an important
magnetic anisotropy leads to spin lifetimes in the ns range. However,
when the magnetic atoms are directly adsorbed on a metal substrate 
the spin lifetime is greatly reduced~\cite{BalashovJAP2010}.
More recent experiments~\cite{K} have measured a lifetime
of 200 fs for the magnetic excitations of individual Fe atoms on Cu (111).
The lifetime decreases by a factor of two when a magnetic field of 12 T is 
applied. The authors explained their findings by the decay of the magnetic
excitation in substrate single-particle excitations or Stoner
excitations where a spin-polarized electron-hole pair is produced.

\subsection{Magnetic adsorbates on metallic substrates}
\label{coupled}

Without a decoupling layer, a magnetic adsorbate on a metallic substrate
is subject to a stronger hybridization and to a direct interaction with
the continuum of electronic excitations of the metal. As we just saw,
the first consequence is the shortened lifetime of adsorbate excitations
due to the large probability of quenching the adsorbate excitation by
exciting the metal substrate.  The short lifetimes lead to an increased
broadening of spectral features. Excitations are poorly resolved in
energy and IETS in general becomes harder to detect on a metal surface.

\begin{figure}
\centering \includegraphics[width=0.8\textwidth]{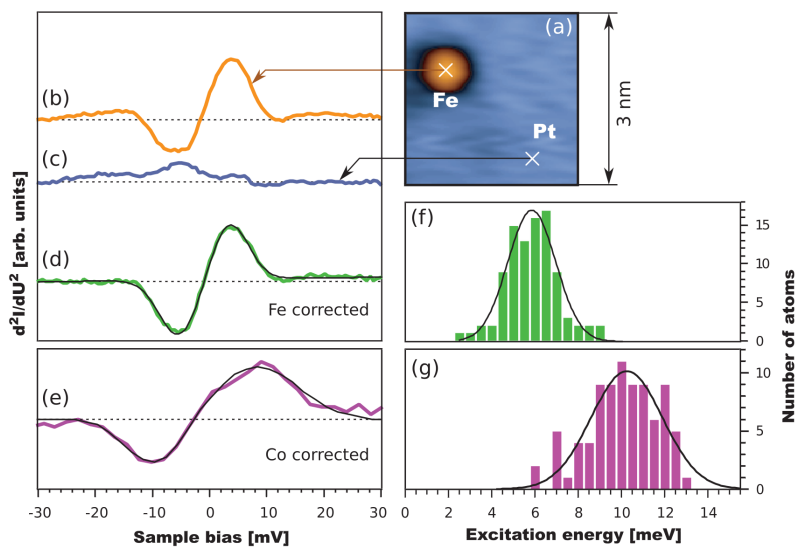}
\caption{\label{FePt111} Magnetic IETS for a Fe atom adsorbed on a Pt
(111) substrate\textcolor{red}{~\cite{BalashovPRL2009}}. (a) STM constant current image of a single Fe atom,
(b) the second derivative of the current with respect to bias or
IETS, (c) IETS of the bare Pt surface, (d) difference between (b) and (c).
The same type of evaluation for a Co atom is shown in (e). (f) and (g)
are the distributions of the measured energies for Fe and Co atoms
respectively. Reprinted with permission 
from reference~\cite{BalashovPRL2009}. Copyright 2009 by the American
Physical Society.} 
\end{figure}

Balashov and coworkers~\cite{BalashovPRL2009} succeeded to measure
magnetic IETS on single Fe and Co adatoms on Pt (111). The
inelastic signals were very small compared to the decoupled
measurements shown above. In order to detect them, Balashov and
coworkers~\cite{BalashovPRL2009} used the second derivative of the
current with respect to bias, Fig.~\ref{FePt111}. This is in stark
contrast to the measurements of Fe, Mn and Co on CuN/Cu (100), see
Section~\ref{decoupled},
 where the analysis of the first derivative was so clear that
the use of the second derivative was not needed. The broadening
caused by the coupling to the Pt (111) continuum of excitations
is very large and of the order of the excitation energy itself,
rendering the identification of the second derivative also
difficult. Nevertheless, a careful statistical analysis permitted
the authors to identify the excitation energies.  

An extra difficulty in these experiments was the  assignment of the
measured peaks to magnetic excitations. The authors achieved it by excluding all other
possible source of the detected peaks. 
The authors claimed that the peaks could not be \textcolor{red}{assigned} 
to collective excitations
such as surface plasmons because they \textcolor{red}{would then} lie in the eV range. Vibrations are in the
same energy range but the authors could not identify any with the measured peaks
by doing a simple harmonic-approximation calculation within density functional theory.
Finally, Kondo physics was left out of the possible causes because Fe and Co are known
to give no Kondo peaks on Pt (111).

The probability of producing the
magnetic excitation by the tunneling current was estimated to
be in the range of 2\% for both Fe and Co on Pt (111). The estimation was \textcolor{red}{obtained} 
by
integrating the area of the second derivative peaks. However, due to the large broadening
it is likely that part of the area is lost by overlapping of the negative and
positive bias peaks, and hence 2\% can be a very low lower limit of the inelastic
efficiency. This IETS efficiency is typical of vibrational excitations~\cite{Hiromu,Komeda}
and hence it is in strong contrast to the efficiencies measured for magnetic excitations of
partially decoupled adsorbates, section~\ref{decoupled}, because they easily were larger than 100\%.
In these systems, several magnetic excitations were also easily detected. On a metal
substrate this is probably impossible because the difference in energies are much
smaller than the intrinsic broadening of the excitations.

Khajetoorians {\em et al}~\cite{KhajetooriansPRL2011}
have also measured the magnetic IETS of Fe adatoms, this time on Cu (111). In this case,
the inelastic signal is directly seen in the conductance measurements. Two symmetric
steps at negative and positive bias appear in the conductance. The step thresholds
shift with the applied magnetic field, following an expected Zeeman splitting, section~\ref{decoupled}, showing   unambiguously that these are magnetic excitations.
The measured gyromagnetic factor $g$ is 2.1, suggesting that the Fe adatom cannot be
described by a pure spin. The step heights are 5\%, again very small compared
to the measured ones in the decoupled case, section~\ref{decoupled}.

As in the experiments by Balashov and coworkers~\cite{BalashovPRL2009}, the broadening
is very large. From the step broadening Khajetoorians {\em et al}~\cite{KhajetooriansPRL2011}
estimate a lifetime of 200 fs, orders of magnitude smaller than
the decoupled case, section~\ref{decoupled}.

Khajetoorians and coworkers~\cite{KhajetooriansPRL2011}
also measured the adatom magnetic moment by using the single-atom magnetization curve (SAMC)
technique~\cite{MeierNature2008}. They used an out-of-plane magnetized STM tip
and measured the relative change in conductance with applied magnetic field. From these
measurements they obtained that the Fe magnetic moment was 3.4 $\mu_B$, quite different
from the 4.0 $\mu_B$ deduced in the IETS experiments of Fe on CuN / Cu(100)~\cite{HeinrichScience2007}.
The value of the Fe magnetic moment is attributed to a consequence of the strong hybridization of
the adatom with the metallic substrate. This led Khajetoorians and coworkers~\cite{KhajetooriansPRL2011}
 to rationalize their findings in terms of an itinerant spin model. This model shows that
the lifetime limiting process of the magnetic excitations is their decay into electron-hole
pairs, namely, Stoner excitations of the itinerant electron gas. This decay
mechanism has also been modelled by Novaes and coworkers~\cite{NovaesPRB2010} albeit
with a different formalism as we will review in the theory section.

\subsection{Magnetic adsorbates on semiconducting substrates}
\label{semicon}

A scenario for spin excitation different from the two seen above (magnetic impurities weakly coupled or
strongly coupled  to the substrate) is the case of magnetic
impurities on a semiconducting surface. Despite the strong chemical interaction,
the absence of conduction states in the semiconductor gap should lead to
a strong charge localization on the impurity. In order to avoid this and
to have some current, the substrates are weakly doped so that the system corresponds more to
a decoupled one instead of a strongly coupled one.

Khajetoorians {\em et al}~\cite{K} have performed
magnetic IETS experiments on Fe adatoms on indium antimonide, InSb (110). The samples
are n-doped and the presence of a small Fe density leads to an accumulation
layer on the surface and to the appearance of a two-dimensional electron gas (2DEG). This
electron gas is characterized by only two electronic bands, one starting at -80 meV
and the other one at -25 meV from the Fermi energy. Hence, in the present
case, there is something like a surface metallic structure permitting the
passage of the tunneling current from the STM tip.

The magnetic IETS are performed ontop of the Fe adatom and two clear
steps, each amounting to 25\% of the elastic conductance (hence 1/3 of
the conductance is inelastic in the present case) at $V\approx \pm 0.5$
mV and $V\approx \pm 1.5$ mV. This very small energy scale comes from
the small MAE as the authors found by fitting Eq.~(\ref{anisotropy})
to $D = -1.4$ meV and $E=0.22$ meV.

In order to rationalize their experimental findings, the authors
used the theory by Lorente and Gauyacq~\cite{Lorente_PRL2009}, see
section~\ref{theory}, with the assumption  that the electron transmission
takes place through a unique electron-adatom spin coupling scheme. By
analyzing the Landau levels of the 2DEG, they realized that the Fe
adatom served as a spin filter and that indeed, only majority spins
were contributing to the electron transmission (the same feature as that
found in Ref.~\cite{Lorente_PRL2009} for Fe on CuN/Cu(100)).

From their analysis, the authors conclude that the Fe spin on InSb (110) is
a $S=1$ system. This is clearly compatible with the three
levels appearing from anisotropy split levels of $S=1$. 
Furthermore, DFT calculations~\cite{K} show that
the Fe atomic configuration on the surface is rather $3d^84s^0$ than
the free atom one $3d^64s^2$. The crystal field from the surface leads
to the single population of two field-split $d$ levels creating a
$S=1$ system in opposition to the $S=2$ free atom as given by Hund's rules.

The 2DEG presents a rich Landau level structure when a magnetic field
is applied.  Analysis of the conductance of the Landau level structure
is sucessfully exploited by the authors to measure the expectation value
of the spin-component along the applied magnetic field. In particular,
these measurement allowed the authors to identify the orientation of the
easy axis and determined the anistropy case unequivocally. Furthermore,
the Landau levels leads to a spin splitting in the tunneling electrons,
hence creating a spin-polarized electron source. Thanks to this spin
polarization the authors could determine the spin-filtering effect of
the Fe adatom.

This set of experiments is a compelling breakthrough in the use of magnetic
impurities in semiconductors, particularly having in mind the possible
application in spintronic devices that currently use semiconducting materials.

\subsection{Collective excitations: magnetic surfaces}
\label{magnon}

The excitation of localized spins partially decoupled from
a metallic substrate, section~\ref{decoupled}, is very
effective and yields large changes in conductance that are easily
measured in an IETS experiment.
As the coupling with the continuum of electronic excitations
of the substrate is increased, the lifetime of the magnetic excitations
is reduced and the magnetic IETS becomes a more difficult technique
with smaller changes in conductance and very broadened features, section~\ref{coupled}.
However, there are other types of magnetic excitations  than
localized spins in an adatom. A magnetic substrate itself can
have excitations, the above mentioned Stoner excitations
are single-particle excitations. There are also collective
excitations where the ensemble of spins of a substrate is excited.
Magnetic substrates can hold waves where their spin change
orientation in a concerted manner. These spin waves are called
magnons when quantized for ferromagnetic materials. 

The first measurement of magnons with magnetic IETS was reported by
Balashov {\em et al}~\cite{BalashovPRL2006,BalashovPRB2008}. They used
the second derivative of the current with respect to bias to detect
peaks when using a Fe (100) crystal and a Co thin film on Cu(111). On
Fe (100) they detected a large peak of $\sim 30$-mV width centered at
3.6 mV. They assigned this peak to the magnon excitation of Fe (100)
with a large change of conductance of 27\%.

In order to prove that this peak is indeed of magnetic origin, they
used a spin-polarized tip and a magnetic field. The magnetic field
was used to change the orientation of the Fe monocrystal. They showed
that the peak height changed with the orientation of the Fe monocrystal.
Magnon excitation entails a spin-flip, hence only minority-spin electrons
from the tip are effected. From these facts, the author concluded
that the large peak at 3.6 mV was indeed the magnetic IETS signal
of magnon excitation in Fe (100).

Further experiments on Co thin films showed that the  conductance
 linearly increases with the number of layers in the Co film. 
Using a phase-space argument, they concluded that this linear
increase was a proof that the excitations were indeed of magnonic origin
and not of phononic one since in this last case the excitations
would extend into the Cu (111) substrate.

 Collective excitations are also found on antiferromagnetic substrates,
but they are more complex than the previous magnon excitations. Spin
waves in antiferromagnetic layers of Mn on Cu$_3$Au (001) were measured
by Gao {\em et al.}~\cite{GaoPRL2008}. The second derivative spectra
permitted them to extract the spin wave dispersion relation
which is linear. The lifetime of the spin waves was seen to scale linearly
with energy in agreement with neutron scattering measurements and
theory, giving a compelling evidence for the excitation of 
spin waves in an antiferromagnetic substrate.

\subsection{Collective excitations: chains of magnetic adsorbates}
\label{chains}

\begin{figure}
\centering
\includegraphics[width=0.8\textwidth]{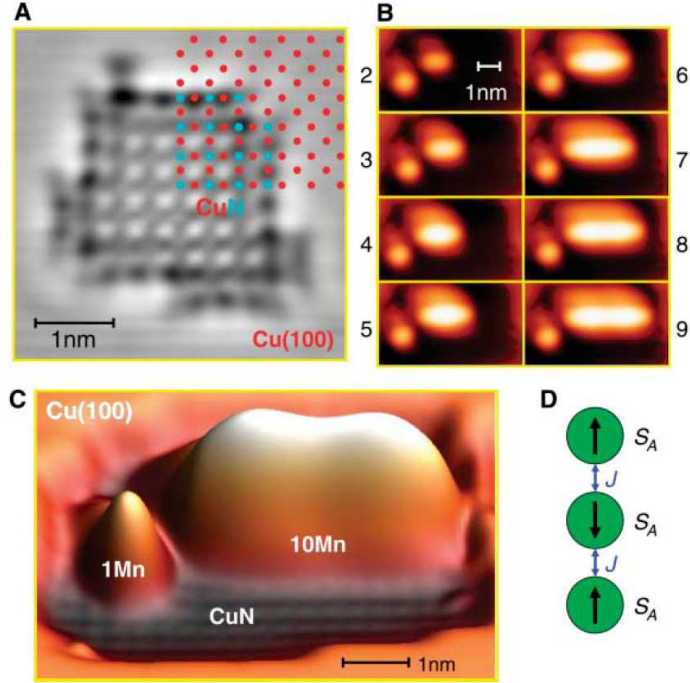}
\caption{\label{Mnchain}
Chain of Mn adatoms on CuN/Cu(100) after Ref.~\cite{HeinrichScience2006}.
In A, an STM constant-current image showing the substrate
structure, B constant-current images of the different Mn chains and C
a comparison of the different elements in the image. D shows
a simple Heisenberg model that accounts for the exchange coupling
among Mn atoms in the chain.  From reference~\cite{HeinrichScience2006}.
Reprinted with permission from AAAS.}
\end{figure}

The atom manipulation capabilities of STM has permitted Hirjibehedin
and coworkers~\cite{HeinrichScience2006} to assemble
finite-size chains of Mn atoms on the CuN on Cu(100) substrate,
Fig.~\ref{Mnchain}. The
magnetic properties of these structures were revealed using
magnetic IETS. The Mn atoms in the chains are antiferromagnetically coupled.
The analysis of the IETS spectra in
terms of a simple Heisenberg chain permitted Hirjibehedin
{\em et al}~\cite{HeinrichScience2006} to find the antiferromagnetic
exchange coupling constant $J$ to be 6.2 meV, varying by around 5\% depending
on the location of the chain in the CuN island. If instead of
ontop Cu sites, the chain was built ontop of N sites, the $J$ value
was halved. The Mn atoms retained their free electron spin
$S=5/2$. 

The excitation process revealed the magnetic structure of
the full chain. This was again proven by comparing the
IETS with the spin states appearing from the diagonalization of
the Heisenberg chain Hamiltonian. Clear differences appear between
chains with an even number of atoms and chains with an odd one.
In the first case the ground state is $S=0$, while odd chains have a finite total spin.
 This leads to a very different spectrum at low
bias, even chains presenting a large elastic gap before
the $S=1$ channel becomes available. 
The triplet character of the first excited state in even chains was
proved by splitting the conductance step in three steps in the
presence of a magnetic field.

The many transitions available as the number of atoms in the
chain changed, led the authors~\cite{HeinrichScience2006}
to the conclusion that there were strict IETS selection rules.
Indeed, the only peaks revealed in IETS where those corresponding
to transitions $\Delta S=0, \pm 1$ and $\Delta S_z = 0, \pm 1$.
These selection rules can be easily explained by
the spin-wave excitation modelled suggested
by Gauyacq and Lorente~\cite{GauyacqPRB2011} where the tunneling
electron first flips the spin of a single Mn atom, and then
the spin-flip propagates through the chain to build the actual excited
state of the full system. 

Recent density functional theory (DFT) calculations~\cite{LichtensteinPRB2009}
show that Mn chains on CuN/Cu(100) present a more complicated magnetic
structure that just a first-neigbors Heisenberg chain. Indeed, the
chain presents some spiral magnetic structure due to the appearance
of non-negligible next-neighbors ferromagnetic couplings. 
From a classical spin Hamiltonian, the authors actually conclude that
the Mn chains present weak ferromagnetism.

Another example of exchange-coupled
magnetic chains was given in Ref.~\cite{ChenPRL2008}. Instead of
an atomically manipulated chain, the authors where studying the magnetic
IETS of several molecular layers of cobalt phthalocyanine. This
molecule replaces the Fe atom of Fig.~\ref{FePc} by a cobalt atom.
The ligand field splitting leads to a d-electron configuration
such that there is only one unpaired electron. Hence the molecule
is a $S=1/2$ system. The magnetic IETS shows however that the
conductance spectrum changes as the number of molecular layers
increases. The change in the IETS can be traced back to the
magnetic interaction between molecules. Assuming a simple exchange
interaction Heisenberg Hamiltonian, the fitting of two and three
layer systems led to an exchange interaction, $J$, of 18 meV between molecules.

\subsection{Polarized electrons: spin torque experiments}
The capacity to controllably switch the magnetization of a magnetic
adsorbate and, also, of a magnetic nanostructure is of great practical
interest. Spintronic devices use spin-polarized currents to change and 
detect magnetic moments and in this way operate. In order to achieve this
the spin-polarized current must exert a spin torque leading to the reversal 
of the local magnetization. Recently, it has been proved
that it is possible to
switch the magnetization of magnetic islands by using 
spin-polarized STM~\cite{KrauseScience2007,HerzogAPL2010}. 

Magnetic IETS is directly connected with spin torque experiments. Above
we cited the experiment by
Loth {\em et al.}~\cite{LothNatPhys2010} where they
 showed that by increasing the
tunneling current they could multiply excite a magnetic adatom. The
current gave them a natural time scale to calibrate the excited
state lifetimes. Now, these excited states actually correspond
to different directions of the atomic magnetization. Hence, by exciting
the adatom, an effective spin torque was transferred from the
tip to the substrate. In order to keep this new magnetization,
the excited state must have a permanent and sizeable population
which leads to large tunneling currents as compared
with the electronic timestep fixed by the excited state lifetime, as
we discussed above.

\subsection{Magnetic IETS and Kondo physics}

The Kondo effect is the screening of the spin of a magnetic impurity
by the conduction electrons of a metallic substrate. The screening
process involves a spin flip of the conduction electron at
no energy cost. It is then an elastic spin flip. The Kondo effect
is then intrinsically related to magnetic IETS. 
The signature of the Kondo effect in an STM experiment is the
appearance of a zero-bias anomaly in the conductance spectra.

The ground state of the magnetic impurity has to be degenerate and
differing in $\Delta S = 0, \pm 1$, $\Delta S_z = 0, \pm 1$ so
that a spin flip takes from one state to the other degenerate state.
However, magnetic anisotropy can lift the degeneracy
preventing the formation of the Kondo state.
Fractionary-spin impurities have at least a Kramers doublet~\cite{Yosida} 
as ground state, hence even in the presence of an important MAE,
the Kondo effect can take place. The Kramers doublet can be destroyed
in the presence of a magnetic field. It is then interesting
to follow the evolution of the Kondo effect with magnetic field.

\begin{figure}
\centering
\includegraphics[width=0.6\textwidth]{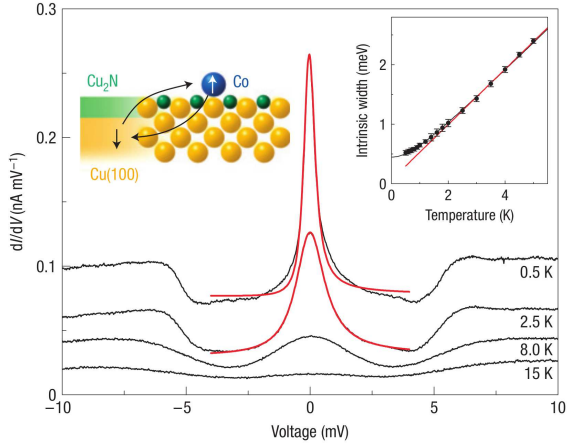}
\caption{\label{CoCuN}
Conductance of an STM junction with a
Co adatom on  CuN/Cu(100).
As the temperature is reduced, the zero-bias anomaly singles out and
the inelastic steps are sharper. The inset shows the fit of the zero-bias
peak width to the behavior of a the Kondo width evolution with temperature
showing that the zero-bias anomaly is indeed a Kondo peak. {Reprinted
by permission from Macmillan Publishers Ltd: Nature Physics~\cite{HeinrichNatPhys2008}, copyright 2008.}
}
\end{figure}

Otte and coworkers~\cite{HeinrichNatPhys2008} have shown the evolution
of the conductance spectra with applied magnetic field in the
case of a Co adatom on CuN/Cu(100), Fig.~\ref{CoCuN}. At zero magnetic field, they
retrieve a Kondo peak at zero bias. Due to the decoupling CuN
layer, the Kondo temperature is very much reduced, from
54 K for the bare Cu (100) surface~\cite{KnorrPRL2002}
to the measured 2.6 K~\cite{HeinrichNatPhys2008}. This makes
possible the observation of the Zeeman splitting of the Kondo
peak at accessible magnetic fields.

In the same conductance spectra, magnetic IETS steps are found,
which permits the determination of the parameters
of Hamiltonian, Eq.~(\ref{anisotropy}), and hence, the
energy structure of the different magnetic states.
The fitting to the spectra yield a hard-axis ($D>0$) anisotropy
that favors the $S_z=\pm 1/2$ as the ground state. An
easy-axis anistropy would have given a $S_z=\pm 3/2$ doublet
that are not linked via a spin-flip impeding the
appearance of a Kondo effect.
As the magnetic field is ramped up, the zero-bias anomaly splits to 
$ 2\Delta$, where $\Delta$ is given by Eq.~(\ref{Zeeman}). 
The rate at which the split peaks separate depends on
the magnetic field direction as dictated by Hamiltonian~(\ref{anisotropy}).

This experiment shows that despite the breaking of the ground state
degeneracy by the applied magnetic field, a remanent Kondo peak is
found at the threshold of the spin-flip transition between $S_z=1/2$
and $S_z=-1/2$ where the threshold is
 $\Delta$, Eq.~(\ref{Zeeman}).  Indeed, at threshold the two states are
degenerate if one considers the total energy including the tunneling
electron: it is the inelastic effect that connects via spin-flip
the $S_z=1/2$ and $S_z=-1/2$. Hence, there is a weak Kondo peak at
threshold. This same behavior is to be expected at any threshold
of spin-flip excitations. Indeed, recent theoretical work assign
the spike-like features at the threshold of some magnetic IETS to
these Kondo peaks~\cite{Hurley:ArXiv2011}.  They have evaluated the
spectra of Fe and Co on CuN/Cu (100) and find some common features
with the experiments~\cite{HeinrichScience2007,HeinrichNatPhys2008}.
However, as shown by Loth and coworkers~\cite{LothNatPhys2010}, the
repeated excitation of the magnetic states as the tunneling current
increases, also produces spikes at threshold. Moreover, Lorente and
Gauyacq~\cite{Lorente_PRL2009} (also see theory section~\ref{theory})
show that not all magnetic excitations in Fe on CuN/Cu(100) are spin
flips. Hence, it is difficult to conclude on the extend of Kondo features
in the above experimental spectra. A theoretical analysis of the
Kondo effect at inelastic thresholds is presented in section~\ref{Kondo}.

However, clear Kondo features at the excitation threshold have
been revealed for the singlet-triplet excitation of carbon
nanotubes by an electron current~\cite{PaaskeNatPhys2006}. These features
have also been detected in the conductance spectra of
a break junction containing C$_{60}$ molecules~\cite{RochNature2008},
and in cobalt complexes where
the Kondo features have been modified via mechanical deformation of
the molecule~\cite{ParksScience2010}.

When an atom of Co is approached by a Fe atom on CuN/Cu(100) the
above Kondo peaks is split by the exchange coupling between the
adatoms~\cite{OttePRL2009}. When a magnetic field is applied
on the sample, the ground and first excited states of Co can
be made degenerate at $B\approx 2$ T in the experimental setup.
This two levels differ in a spin-flip and they become degenerate
at $B\approx 2$ T. As a consequence the full zero-bias
anomaly is restored. The measured conductance spectrum is
basically equal to the single Co one~\cite{OttePRL2009}.

%

\section{Theoretical results}
\label{theory}


\subsection{Computation of magnetic anisotropy in adsorbates (ab initio)}

The magnetic anisotropy of an isolated adsorbate on a surface
is very often described using the effective anisotropy Hamiltonian (\ref{anisotropy}) described in section~\ref{decoupled}  with the D and E anisotropy coefficients.
It assumes the existence of a local
spin, carried by the adsorbate, the direction of which is influenced by
the surroundings. The origin of this effective Hamiltonian is described
in detail in Ref~\cite{Yosida} (see also descriptions of the magnetic
anisotropy using tight-binding approaches in ~\cite{Autes_JPC2006,
Desjonqueres_PRB2007,Bruno_PRB1989}). Briefly, in the case of a
system with light atoms, the spin-orbit interaction can be considered to
be weak, at least much weaker than the electrostatic interactions. In
a calculation without spin-orbit interactions, the ground state
is associated to a configuration of molecular (atomic) orbitals
corresponding to a certain total spin of the adsorbate ($\vec{S}^2$) and
degenerate with respect to the spin projection on the quantization axis
($\hat{S}_z$). Indeed, at this level of approximation, the adsorbate
spin corresponds to the filling of the orbitals and does not interact
with the adsorbate surroundings. The spin-orbit interaction can then be
introduced as a perturbation that couples this ground state with excited
states with different spin and orbital angular momenta. It leads to a
second order correction on the energy, lifting the degeneracy of the
ground state and splitting the different $\hat{S}_z$ states. It can be
stressed that, once the spin-orbit interaction is taken into account,
the adsorbate states correspond to mixings of different spin and
orbital angular momentum states and cannot, in principle, be labelled
only considering $\hat{S}_z$. Though, if the spin-orbit perturbation is
weak\textcolor{red}{, the state mixing is weak, each state is dominated by a single $\hat{S}_z$ value and }  one can keep the $\hat{S}_z$ labelling of the different states as
an approximation and represent the second order energy splitting of the
ground state using Hamiltonian (\ref{anisotropy}). The D and E parameters
then describe the  way the adsorbate magnetic moment orientates with
respect to the substrate and they are obtained by perturbation theory~\cite{Yosida}. If a magnetic field, B, is applied, its action
can be represented by a Zeeman term: $g \mu_B \vec{S} \cdot \vec{B}$, added to
the Hamiltonian (\ref{anisotropy}). $\mu_B$ is the Bohr magneton and $g$
the Land\'e factor. Here again, one assumes that the states coming from
the splitting of the ground state can be described by only considering
$S$ and possible extra mixings are introduced in the effective $g$ factor.

$Ab$ $initio$ computation of the anisotropy coefficients
(D and E coefficients for light adsorbates)  \textcolor{red}{ then implies taking the
 spin-orbit interactions into account}. Several calculations,
based on Density Functional Theory (DFT), have been performed
for 
adsorbates on different substrates that had been studied
experimentally~\cite{LichtensteinPRB2009,Blonski-Hafner2009,ShickPRB2009,Sipr_PRB2010,Shick_PRB2009,Blonski_PRB2010,Barral_PRB2010,Wang2009,Bluegel2011}, a discussion of the
assessment of the involved approximations can be found in Blonski and
Hafner~\cite{Blonski-Hafner2009b}. The DFT-based approach consists in
introducing the spin-orbit as an operator involving the electron spin
and added to the potential felt by the electrons~\cite{Blonski-Hafner2009,SIESTA_SO,Richter2009,BLAHA}. It is thus possible to compute the total energy of the
adsorbate+substrate system for different orientations of adsorbate
magnetic moments, this yields the magnetic anisotropy energy (MAE), or
in other words, the magnetic landscape for the adsorbate. An adjustment
procedure can then be used to extract the D and E coefficients that
reproduce the computed MAE. Evaluation of the magnetic anisotropy
thus involves making differences of total energies to get small energy
terms. Typically for the case of Mn and Fe adsorbates on CuN/Cu(100)
that have been studied experimentally by IETS~\cite{HeinrichScience2007}, the
magnetic anisotropy is in the meV energy range, so, a priori, it requires
high accuracy calculations~\cite{Blonski-Hafner2009,Sipr_PRB2010}. This
difficulty might account for the limited success of the ab initio
evaluation of magnetic anisotropy terms in accurately accounting for
IETS observations.

\subsection{Treatment of inelastic transitions  induced by tunnelling electrons in a local spin }

The experimental observations of magnetic excitations by IETS that could
be interpreted as magnetic transitions of a local spin in invidual
adsorbates prompted a series of theoretical studies, aiming at the
understanding and description of the phenomena at play. The first accounts
were considering one-electron approaches in which the tunnelling electron
is directly inducing the transitions of the local spin.

\subsubsection{$S^2$ theories}
\label{S2}

The first theoretical treatment of inelastic magnetic transitions
accompanied one of the first experimental observations~\cite{HeinrichScience2007}
of magnetic IETS. In IETS, the height of each step in the conductance
corresponds to the strength of an inelastic transition, whereas the
conductance at zero bias yields the elastic conductance. IETS thus
yields a detailed account of the magnetic transitions induced by a
tunnelling electron: the ratio between the zero-bias conductance and
the conductance steps at finite bias yields the relative probability
of elastic and inelastic scattering for a tunnelling electron with
an energy larger  than all inelastic thresholds.  Starting from the
anisotropy Hamiltonian~(\ref{anisotropy}), with a Zeeman term added,
it is  diagonalized in the basis set formed by the $|S,M\rangle$ states,
the eigenstates of $\vec{S}^2$
 and $S_z$, so that the anisotropy eigenstates can be written as:

\begin{equation} 
\left|\phi_{n}\right\rangle = \sum_{M} C_{n,M}
\left|S,M\right\rangle .
\label{phi_n} 
\end{equation} 

The $\phi_n$ eigenstate of the system is associated to the $E_n$
 eigenenergy. The steps in the conductance correspond to transitions from
 the ground state $\phi_1$ to the excited $\phi_j$ states induced by the
 tunnelling electrons. By analogy with neutron scattering, the authors
 considered the $\vec{S}$ operator as the transition operator and showed
 that the experimental ratio between the various inelastic channels could be very
 well accounted for by the ratio between the matrix elements squared of
 $\vec{S}$ between initial, $\phi_1$, and final states, $\phi_j$. This
 appears very clearly in Fig.~\ref{Fe} that shows 
the experimental and calculated tunneling conductance over
a Fe atom on CuN/Cu(100)~\cite{HeinrichScience2007}. The $S^2$ scaling reproduces the ratio between inelastic channels; however, 
 this scaling  makes wrong predictions on the elastic channel and thus
 it cannot account for the overall strength of the magnetic transitions.

An extension of the original $S^2$ theory has been used in  fitting
procedures by Loth et al~\cite{Loth_NJP} and Chilian et al~\cite{Chilian_2011}. It consists in using a
transmission amplitude equal to ($u  +  {\vec{s}} \cdot {\vec{S}}$), with $u$
as an effective adjustable parameter. The $u$ term only contributes
to elastic tunnelling and added to the usual term of the original $S^2$
theory, it allows to fit the elastic channel at the desired level and thus
to bypass the difficulty mentioned above about the original $S^2$ theory
(vanishing $u$), which was unable to account for the elastic transmission.

Rationalization of the $S^2$ scaling
was later provided within perturbation theory
approaches~\cite{FernandezRossier_PRL2009,Fransson2009}. The basic idea
is to introduce in the Hamiltonian describing the  tip-sample system a
tunnelling term including the product of the
electron spin times the adsorbate spin. 
Perturbation theory then allows to compute the current
flowing through the tip and the adsorbate. This approach recovers the
$S^2$ scaling for the excited channels, since the tunnelling term of the
Hamiltonian contains the adsorbate spin as a factor. Its prediction is
typical for a perturbative approach: the probability of the inelastic
transition is proportional to the matrix element of the coupling between
initial and final states. But it cannot make predictions on the elastic
channel. This is not surprising for a process in which the inelastic
transmission is dominating over the elastic one, perturbation theory
cannot be expected to work.

Persson~\cite{Mats_PRL2009} went further in justifying the $S^2$
scaling. He made use of a sudden approximation (impulsive approximation)
to treat tunnelling through the magnetic adsorbate. The magnetic
anisotropy terms are very small and  one can then  assume that they are
not active during electron tunnelling through the adsorbate, which is a
very fast process in the absence of a very long-lived adsorbate-localised
resonance. The electron tunnelling amplitude in the absence of magnetic
anisotropy is expressed using Tersoff-Hamann~\cite{TersoffHamann}
approximation taking into account the spin dependence
of the T-matrix assuming complete spherical symmetry as justified by the
sudden approximation. Then, the elastic and inelastic amplitudes are evaluated
as the matrix element of the Tersoff-Hamann tunnelling amplitude between
the initial and final states of the adsorbates, taking the anisotropy
into account. In this way, the tunnelling current appears as the sum
of two terms, one independent of the adsorbate spin variables and one
proportional to the squared matrix element of the $\vec{S}$ operator
between initial and final states. This recovers the $S^2$ scaling of the
inelastic channel probabilities without resorting to perturbation theory,
but in the absence of a quantitative evaluation of the ratio between
the two terms in the tunnelling current, it cannot make a quantitative
prediction for the magnitude of the inelastic current with respect to
the elastic current. Though, one can stress that the description of the relative importance
of the various inelastic channels, irrespective of the elastic channel, by the so -called $S^2$
scaling is thus \textcolor{red}{fully} justified.

\subsubsection{Strong coupling approach}
\label{Strong}

A strong coupling theory has  been developed for the magnetic IETS
transitions; it is a one-electron approach that explicitly introduces the
spin symmetries of the problem and leads to a simple formulation, allowing
quantitative predictions on the strength of the elastic and inelastic
channels \cite{Lorente_PRL2009,Novaes_FePc}.  The effect of the anisotropy Hamiltonian (\ref{anisotropy}) is treated in the
sudden approximation, i.e. one defines a tunnelling amplitude between the
tip and the substrate, noted  $T_{Tip\rightarrow Sub}$ (and an equivalent
one for the reverse tunnelling) without the anisotropy Hamiltonian
terms taken into account.  It corresponds to  the tunnelling electron
scattering from the adsorbate and it depends on the electron energy. In
the absence of magnetic anisotropy (i.e. without spin-orbit interaction),
it depends  on the spin coupling between the tunnelling electron
and the adsorbate, via the exchange interaction. Thus, it can be written
in a diagonal form if we consider $\vec{S}_T$, the total spin of the
(electron + adsorbate) system. Defining $\left|S_T,M_T\right\rangle$
as the eigenfunctions of $\vec{S}_T^2$  and $S_{T,z}$  (if $S$ is
the adsorbate spin, then $S_T = S \pm \frac{1}{2}$), we can write
formally the scattering $T_{Tip\rightarrow Sub}$  matrix (in the
absence of magnetic anisotropy) as: 

\begin{equation} 
T_{Tip\rightarrow Sub}= \sum_{S_T,M_T} \left|S_T,M_T\right\rangle T_{Tip\rightarrow
Sub}^{S_T} \left\langle S_T,M_T\right|.  
\label{Tmatrix} 
\end{equation}
$T_{Tip\rightarrow Sub}^{S_T}$ is a complex number; it depends  on
the electron energy. However, \textcolor{red}{ the total span of bias
that is considered in magnetic IETS is very small, so that $T_{Tip\rightarrow Sub}^{S_T}$ }can be considered to be constant.

 In the sudden approximation, the tunnelling amplitude (in the presence
of magnetic anisotropy) is written as the matrix element of the
$T_{Tip\rightarrow Sub}$ amplitude between the initial and final states
of the tunnelling process. These states are written as $|\frac{1}{2}, m;
\phi_n \rangle $ where the first part concerns the tunnelling electron
(the electron spin is 1/2 and $m$ is the projection of the tunnelling
electron spin on the quantization axis) and the second part concerns
the local spin of the adsorbate.  One then obtains the amplitude,
$AMP_{m,n\rightarrow m',n'}$, for a tunnelling electron induced
transition from $\phi_n$ to $\phi_{n'}$, while the tunnelling electron
spin projection changes from $m$ to $m'$ as:

\begin{eqnarray}
&AMP_{m,n\rightarrow m',n'} = 
\sum_{S_T} \; T_{Tip\rightarrow Sub}^{S_T} & \nonumber \\
&\times
\sum_{M_T} \langle \frac{1}{2}, m'; \phi_n' | S_T,M_T \rangle
\langle S_T,M_T | \frac{1}{2}, m; \phi_n \rangle&
\label{amp}
\end{eqnarray}

One can see that the tunnelling amplitudes associated to the two $S_T$ symmetries 
are interfering. 
Equation~(\ref{amp})
 yields the transition amplitudes in the fully determined case, when
the projection of the spin of the tunnelling electron is registered
in both the initial and final states. Implicitly, it has been assumed
above that the tunnelling electron quantization axis is the z-axis
of the adsorbate magnetic anisotropy; situations with different
quantization axis for the adsorbate and the tunnelling electron can be
easily handled with an expression similar to Eq.~(\ref{amp}).  We can
now define the probability, $P_{n\rightarrow n'}$, for transitions
from $\phi_n$ to $\phi_{n'}$ induced by unpolarized tunnelling
electrons by summing incoherently over the distinguishable channels,
as given by the tunneling electron's spin orientation $m$
for the initial channel and  $m'$ for the final one:

\begin{eqnarray} 
P_{n\rightarrow n'} &=& \frac{1}{2} \sum_{m,m'} |
AMP_{m,n\rightarrow m',n'}|^2 \nonumber \\ &=& \frac{1}{2} \sum_{m,m'}
| \sum_{S_T} T_{Tip\rightarrow Sub}^{S_T}   \\ &\times& \sum_{M_T}
\langle \frac{1}{2}, m'; \phi_n' | S_T,M_T \rangle \langle S_T,M_T |
\frac{1}{2}, m; \phi_n \rangle |^2. \nonumber 
\label{prob} 
\end{eqnarray}

The  eigenstates of the total spin,$|S_T,M_T\rangle$, can be expressed explicitly
as expansions over products of the adsorbate and electron spin states:
\begin{equation}
|S_T,M_T\rangle = \sum_m CG_{S_T,M_T,m} |S,M=M_T-m\rangle |\frac{1}{2}, m\rangle
\label{SM}
\end{equation}
where $ |S,M\rangle$ states
correspond to the adsorbate spin states and $|\frac{1}{2}, m\rangle$
to the tunnelling electron spin. The CG are Clebsch-Gordan coefficients. 
Combining (\ref{phi_n}) and (\ref{SM}) we
can express the total spin states as expansions over  products
of adsorbate magnetic anisotropy states and tunnelling electron spin:
\begin{equation}
        \left|j\right\rangle  =\left|S_{T},M_{T}\right\rangle =
        \sum_{n,m} A_{j,n,m} \left|\phi_{n}\right\rangle        \left|1/2,m\right\rangle 
\label{Ajnm}
\end{equation}

The transition probability (\ref{prob}) then becomes:
\begin{equation}
P_{n\rightarrow n'}  
= \frac{1}{2} \sum_{m,m'} | \sum_{S_T} T_{Tip\rightarrow Sub}^{S_T}   
\sum_{M_T} A_{j,n,m} A^*_{j,n',m'}|^2.
\label{P2}
\end{equation}

In several cases, a DFT study revealed that $T_{Tip\rightarrow Sub}^{S_T}$
  tunnelling amplitude is dominated by a single symmetry,
  $S_T$. This was found in the case of Mn and Fe adsorbates on
  CuN/Cu(100)\cite{Lorente_PRL2009} as well as in the case of FePc (iron
  phthalocyanine) on CuO/Cu(110)\cite{Novaes_FePc}. It can be simply
  interpreted as the tunnelling through the adsorbate being dominated by
  a single orbital in the small energy range scanned in magnetic IETS.
  In that case,
the probability (\ref{P2}) further simplifies into: 

\begin{equation}
P_{n\rightarrow n'} = \frac{1}{2} | T_{Tip\rightarrow Sub}^{S_T}|^2
\sum_{m,m'} | \sum_{M_T} A_{j,n,m} A^*_{j,n',m'}|^2.  
\label{P3}
\end{equation} 

This result, used in Ref.~\cite{Lorente_PRL2009}, is very
simple, the electronic part of the tunnelling (the $T_{Tip\rightarrow
Sub}^{S_T}$ amplitude) is factored out and the probabilities for the
different channels are simply proportional to spin-coupling coefficients
corresponding either to the magnetic anisotropy or to the coupling
between electron and adsorbate spins (the coefficients are products of
the diagonalization expansion coefficients in Eq.~ (\ref{phi_n})
 and Clebsch-Gordan coefficients).

 The conductance $dI/dV$ as a function of the STM bias, $V$, can then be
written as: 

\begin{equation} 
\frac{dI}{dV}=C_0\frac{\sum_n \Theta(V-EX_n)
\sum_{m,m'} |\sum_j A_{j,1,m} A^*_{j,n,m'}|^2}{ \sum_n \sum_{m,m'}|\sum_j
A_{j,1,m} A^*_{j,n,m'}|^2}.  
\label{conductance} 
\end{equation}

Expression~(\ref{conductance}) corresponds to the conductance for
the system being initially in the ground state $n=1$. The sum over $n$
extends over all the $| \phi_n \rangle $ states, including the ground
state, so that the above conductance takes
 {all contributions, elastic and inelastic,} into account. $EX_n$ is
  the excitation energy
of the magnetic level $n$, corresponding to the eigenvalue difference
between the final, $| \phi_n \rangle $, and initial $| \phi_1 \rangle $
states. The Heavyside function, $\Theta$, takes care of the opening of the
inelastic channels at zero temperature.  $ C_0$ is the total conductance
corresponding to the  transmission amplitude $T_{Tip \rightarrow
Sub}^{S_T}$ and is then a magnetism-independent conductance. Since we
only consider a limited $V$ range, defined by the magnetic excitation
energies, $C_0$ can be considered as constant in the relevant $V$-range.
$C_0$ is equal to the conductance of the system for biases larger than all
the inelastic magnetic thresholds.  Expression~(\ref{conductance}) corresponds to
the case where only one $S_T$ value actually contributes to tunnelling
so that the sum over $j$ is restricted to the corresponding $M_T$
values. If the two $S_T$ symmetries contribute to tunnelling, a more
general expression derived from Eq.~(\ref{P2}) has to be used.

At this point, one can stress the main characteristics of the magnetic
tunnelling described in Eq.(\ref{P2} and \ref{conductance}). It reduces to sharing the electron flux \textcolor{red}{associated to } a global
conductance, $C_0$, independent of the magnetic anisotropy, among the
various anisotropy states.  The inelastic transmission probability thus
does not appear as a squared matrix element of a coupling but as the
result of the coupling and decoupling of the electron spin with that of
the adsorbate. One can see the tunnelling through the adsorbate as a spin
filter that selects a specific spin-coupling between the electron and the
adsorbate (the $S_T$ symmetry); the probability of a certain $n$ to $n'$
transition is obtained by projecting the initial state(adsorbate $n$ and
spin of the electron $m$) on the spin-filter states at the beginning and
by projecting the spin-filter states on the final states (adsorbate $n'$
and spin of the electron $m'$) at the end of tunnelling.

One can also stress that if one is only interested in the ratio between
the various elastic and inelastic channels (the information yielded by
magnetic IETS) and not so much by the absolute value of the  tunnelling
current, then only a limited input is needed to describe the process:
the anisotropy constants, the spin of the adsorbate, $S$, and that of
the tunnelling, $S_T$. Because of this, the above formalism can also be
used to analyze experimental data in the absence of an {\em ab~initio} study:
the spectroscopic information (energy positions of the conductance steps)
can be used to model the magnetic anisotropy and the step height can be
used to yield the involved spin-symmetries. This strategy has been very
efficiently applied by Khajetoorians et al~\cite{K} in
their study of magnetic transitions in Fe atoms on an InSb(110) surface.

\subsubsection{Iron phthalocyanine molecule adsorbed on CuO/Cu(110)}

\begin{figure}
\centering
\includegraphics[width=0.7\textwidth]{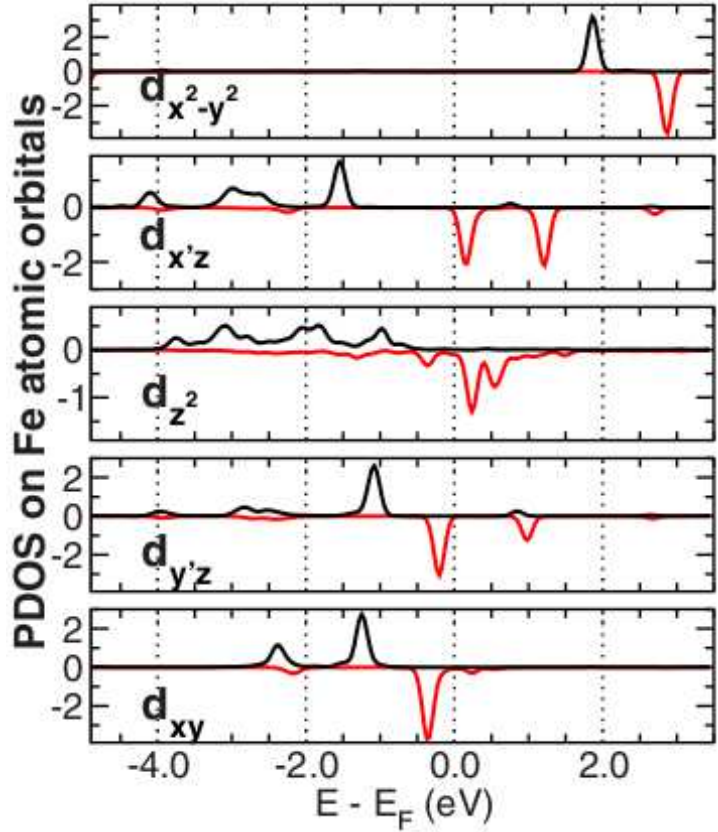}
\caption{\label{pdos_fepc}
{Projected density of states (PDOS) on the Fe d-atomic orbitals \textcolor{red}{in the case of an Fe-Phthalocyanine molecule adsorbed on a CuO/Cu(100) surface}. 
For all the curves shown here, the positive (black) corresponds to the majority spin, and the negative (red) to the minority spin. The $d$ orbitals are classified 
according to the Cartesian axes that contain the N-Cu-N axis of the molecule ($x$,$y$,$z$), 
or with respect to the surface directions: $x'$ for the [$1\bar{1}0$] and $y'$ for the [$001$] directions. 
The $z$ axis is the same for both reference frames.} \textcolor{red}{ Taken from reference~\cite{Novaes_FePc}.}
}
\end{figure}

\begin{figure}
\centering
\includegraphics[width=0.4\textwidth]{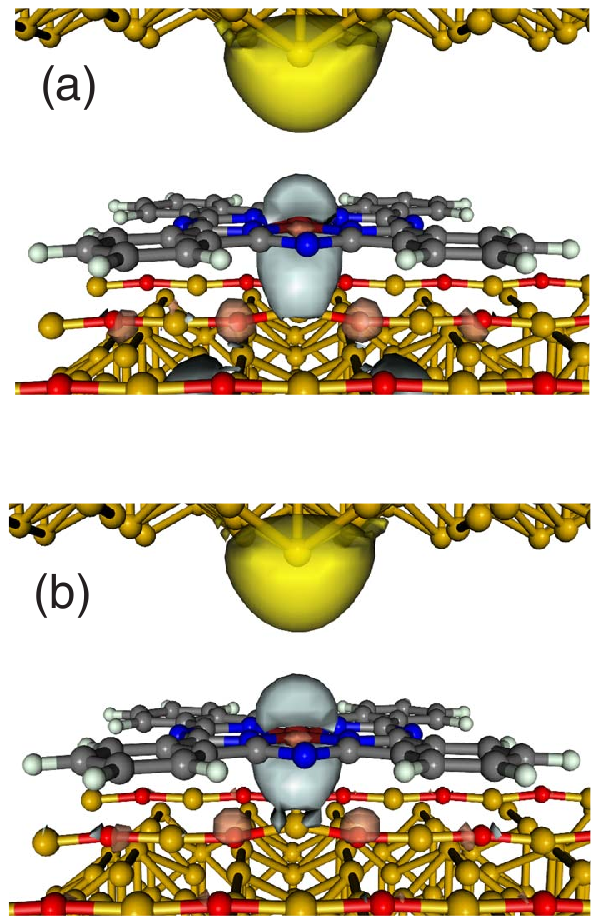}
\caption{\label{teig_fepc}
Transmission between a Fe-phthalocyanine molecule adsorbed on a CuO/Cu(100) surface and an STM tip modeled as an extra Cu atom on Cu(100).
Transmission eigenchannel corresponding to the largest transmission amplitude for (a) $E=E_F$ where $E_F$ is the 
system's Fermi energy and (b) $E=E_F+0.2$ eV. Light gray (light
pink) color corresponds to the positive (negative) imaginary part of
the eigenchannel amplitude coming from the STM tip. In gold
color, the positive real part. Note that the isosurfaces were chosen
differently in (a) and (b) because of the large difference in transmission
probability. The transmission channel exhibits a strong $a_{1g}(d_{z^2})$ character around the Fe atom.  Taken from reference~\cite{Novaes_FePc}.
}
\end{figure}

The case of iron phthalocyanine (FePc) molecules adsorbed on
CuO/Cu(110) has been studied in detail using the strong coupling
approach~\cite{Novaes_FePc}, following the experimental work of
Tsukahara et al~\cite{TsukaharaPRL2009}. In a first step, the structure
of the FePc/CuO/Cu(110) system has been studied ab initio using the
SIESTA package~\cite{RefSIESTA} (only the higher symmetry $\beta$
adsorption geometry has been studied). The local spin of the system
is localized on the central Fe atom of the molecule. The s-electrons
of Fe are transferred to the rest of the molecule and the Fe  retains
its $3d^6$ electronic structure. In the free FePc, the field around
the Fe atom is very strong and of $D_{4h}$ symmetry . Together with
the adsorption effect, this results in a large splitting of the $3d$
manifold (see in~\cite{Dale_1968a, Lu1997, Liao2001,Marom2009} and references therein for a discussion of the electronic structure of free FePc molecules). Real d orbitals are best adapted to discuss the $D_{4h}$
symmetry and we keep these for the orbital notation, even in the case of
the adsorbed molecule. The $d_{x^2-y^2}$ orbital is very high in energy
and unoccupied, whereas the $d_{xy}$ is very low in energy and fully
occupied. The spin structure of the adsorbed molecule then correspond to
four electrons occupying the $d_{xz}$, $d_{zy}$ and $d_{z^2}$ orbitals
(note that upon adsorption the $d_{xz}$ and  $d_{zy}$ are not degenerate
anymore). Following Hund's rule, the molecule is then predicted to have a spin $S$ equal to 1,
as observed experimentally. The projected density of states (PDOS) of the
molecule for the various d-orbitals of Fe are shown in Fig.~\ref{pdos_fepc} and we can conclude that the Fe in the adsorbed molecule
can be attributed the effective $d_{xy}^2d_{y'z}^2d_{x'z}d_{z^2}$
configuration of molecular orbitals. In a second step, the conductance
between a tip and the substrate through the central Fe atom has been
computed using the {\sc  TRANSIESTA} package \textcolor{red}{for non-equilibrium transport calculations ~\cite{RefTRANSIESTA,Novaes2006}}.  In the
absence of spin-orbit couplings, this package allows to compute the
transmission amplitude between tip and substrate, free from the magnetic
anisotropy effects, i.e. exactly the transmission amplitude that is
needed as input in the sudden approximation involved in the strong
coupling approach. Furthermore, this package allows to determine the
eigenchannels for the transmission~\cite{Frederiksen2007,Paulsson2007}.
Figure \ref{teig_fepc} presents the transmission eigenchannel with the highest
transmission for two different energies: the Fermi energy and 0.2
eV above. Besides the big lobe centred around the tip, that is to be
expected for a transmission eigenchannel, it exhibits a double lobe
structure around the Fe atom, strongly suggestive of a $d_{z^2}$ Fe
atomic orbital perturbed by the surroundings. In addition, the same
orbital shape is found for the Kohn-Sham orbital found around 0.2
eV above the Fermi level. From this, we can then conclude that tunnelling
through the Fe atom dominantly involves the Fe-$d_{z^2}$ orbital and so,
that tunnelling can be  described by the two effective configurations
: $d_{xy}^2d_{y'z}^2d_{x'z}$  and $d_{xy}^2d_{y'z}^2d_{x'z}d_{z^2}^2$
for holes and electrons, respectively. Both configurations are doublet,
hence the intermediate tunnelling symmetry ($S_T$
in the strong coupling approach) is equal to 1/2.

\begin{figure}
\centering
\includegraphics[width=0.4\textwidth]{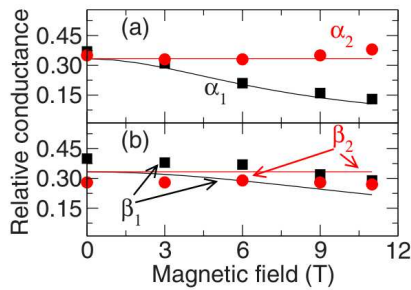}
\caption{\label{relat_cond_fepc}
{\textcolor{red}{Magnetic IETS for an STM tip located above the Fe atom of an Fe-phthalocyanine molecule adsorbed on a CuO/Cu(100) surface.}
Relative inelastic step heights in the conductance
as a function of the magnetic field $B$, for (a) $\alpha$ and (b) $\beta$ configurations. $\alpha_1$ and $\alpha_2$ 
refer to the first and second excitation steps for the $\alpha$ configuration, respectively. Analogously, $\beta_1$ and $\beta_2$ 
refer to the first and second excitation steps for the $\beta$ configuration.
The experimental data points are represented with black squares for
the first excitation and as red circles for the second one and are
taken from the supplemental material of~\cite{TsukaharaPRL2009}. \textcolor{red}{The  theoretical results (red and black lines) are obtained in the strong coupling approach. Taken from reference~\cite{Novaes_FePc}.}
}
}
\end{figure}

We then have all the required inputs to compute the magnetic transitions
induced by tunnelling electrons (if we assume the magnetic anisotropy
Hamiltonian known from experiment). The corresponding results are shown in
 Fig.~\ref{relat_cond_fepc} which shows the relative magnitude of the
various conductances, elastic and inelastic, as function of the applied
magnetic field, B. Both $\alpha$ and $\beta$ geometries are shown; it
was assumed that the spin symmetries obtained ab initio in the case
of the  $\beta$ geometry are also valid for $\alpha$ geometry. The
strong coupling results reproduce extremely well the experimental
results \textcolor{red}{(Note that in Fig.~\ref{relat_cond_fepc}, the step heights are normalised so that the sum of elastic and inelastic conductances is equal to 1; the excellent agreement seen in Fig.\ref{relat_cond_fepc} for the inelastic conductance is thus implying an excellent agreement for the inelastic/elastic ratio)}. In particular, for B = 0,  the three channels are equivalent,
thus corresponding to an inelastic conductance twice larger than the
elastic one. The B-variation of the relative
magnitude of the various channels is also very well accounted for in
the strong coupling approach. Actually this variation corresponds to the
change of the magnetic structure of FePc from a structure dominated by
the magnetic anisotropy imposed by the substrate at B = 0 to a regime
dominated by the Zeeman effect at large B. The variation seen in
Fig.~\ref{relat_cond_fepc} is thus the effect of the decoupling of the
magnetic anisotropy by the B-field (this decoupling occurs earlier in the
$\alpha$ geometry for which the transverse anisotropy, E, is weaker). We
can see on this example the capability of magnetic IETS to emphasize
the changes in the magnetic structure of an adsorbed system.

We can use the case of FePc to discuss and illustrate the selection
rule in magnetic excitations. Since the transitions are induced by a
spin-1/2 particle, one could say that the only possible transitions are
associated with the ${\Delta}M$ = 0,$\pm$1 selection rule (only the latter
two are associated with a spin-flip of the exciting electron). However,
such a selection rule only applies in systems where  $M$, the projection
of the spin of the adsorbate on the quantisation axis, is a good quantum
number. In the case of adsorbed FePc, at B = 0, the adsorbate spin is $S$
= 1,  the ground and the first excited states of the system are mixtures
of the $M$ = +1 and $M$ = -1 states, whereas the highest state is the
$M=0$ state (see details in~\cite{TsukaharaPRL2009}); the ${\Delta}M$
selection rule then does not apply. Though, one can notice that excitation from the ground
state to the first excited state is  a non-spin-flip transition for the
electron, whereas the excitation of the second state is spin-flip. For large
B fields, the Zeeman effect dominates and the various states become
eigenstates of $M_z$. Excitation of the ground state ($M$ = -1 state)
then only occurs towards the $M$ = 0 state (the second excited state
in the B-range spanned in Fig.~\ref{relat_cond_fepc}) and one can see that the magnetic
excitation is indeed dominated by the excitation toward the second state
in Fig.~\ref{relat_cond_fepc}, as a consequence of ${\Delta}M$ = 0,$\pm$1 selection rule;
this dominant process is of spin-flip character.

\subsubsection{Fe adsorbates on CuN/Cu(100) surfaces}

\begin{figure}
\centering
\includegraphics[width=0.6\textwidth]{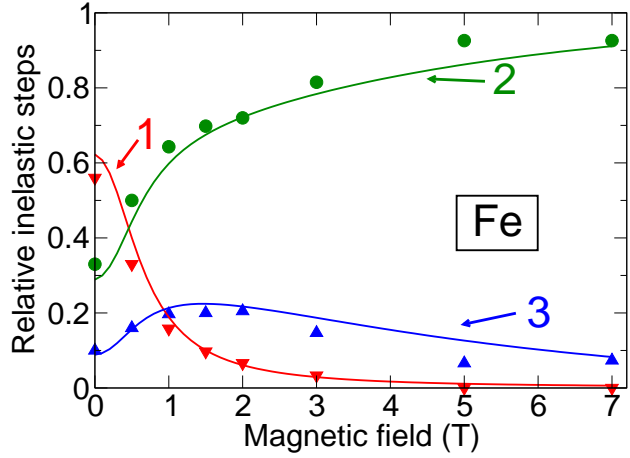}
\caption{\label{figure_Fe-Th}
Relative heights of the inelastic steps in the conductance of an Fe atom on CuN/Cu(100), i.e. relative weights of the inelastic transmission probabilities, as functions of the applied magnetic field, $B$. Symbols: experimental data from ~\cite{HeinrichScience2007} and full lines: theoretical results obtained with the strong coupling approach~\cite{Lorente_PRL2009}. The numbers (1-3) label the excited magnetic Fe states following their excitation energy order. The highest state (number 4) is only very weakly excited and is not included in the figure. \textcolor{red}{ Taken from reference~\cite{Lorente_PRL2009}.}
} 
\end{figure}

The case of Fe adsorbates on a CuN/Cu(100) substrate, which has been studied experimentally~\cite{HeinrichScience2007} and theoretically~\cite{Lorente_PRL2009}, is also illustrative of the characteristics of the magnetic excitation process. As discussed in section 2, the adsorbed Fe atom is associated to a local spin $S=$ 2. The anisotropy brought by the substrate and an applied magnetic field splits the $S=$ 2 manifold into five states. Analysis of the tunnelling process by a DFT study shows that tunnelling is associated with a $S_T$ = 5/2 symmetry~\cite{Lorente_PRL2009}.  Figure~\ref{figure_Fe-Th} presents the relative heights of the inelastic steps in the conductance of an Fe atom on CuN/Cu(100), i.e. the relative weights of the inelastic transmission probabilities, as functions of the applied magnetic field, $B$. The numbers (1-3) label the excited magnetic Fe states following the order of their excitation energy. The theoretical results are seen to reproduce extremely well the experimental data and in particular their variation with the applied $B$ field and the absence of excitation for one of the excited states. The highest level (number 4) is found to be only weakly excited in the theoretical approach and not observed experimentally, it has not been included in the figure.

 The variation of the excitation probabilities with $B$ reflects the change of magnetic structure of the system. At $B = 0$, the two lowest states correspond to the mixing of the two $|S_z|=$ 2 states by the transverse magnetic anisotropy ($E$ term in Eq.~(\ref{anisotropy}), \textcolor{red}{see also Table 1}). In this case, the strong 0-1 excitation is induced by the coupling-decoupling of the transverse  anisotropy by the collision electron (initially the $S_z= \pm2$ are coupled by $E$, they couple independently to the tunnelling electron spin during the collision and are coupled again together by the $E$ anisotropy at the end of the collision) ; such an excitation process is not associated to a spin-flip of the electron. In contrast, still at $B$ = 0, the 0-2 and 0-3 transitions that involves transitions between $S_z= \pm $ 2 and $S_z= \pm$ 1 states are associated with a spin-flip of the tunnelling electron. 
At very large magnetic $B$ fields, the system shifts to a Zeeman structure where the anisotropy effect can be neglected and where each state is associated to a given value of $S_z$. Then the only transitions inside the Fe $S =$ 2 manifold are of spin flip type, they involve the $\Delta S_z= \pm1$ selection rule. This aspect appears clearly in Fig.~\ref{figure_Fe-Th}, where at large $B$, only the 0-2 transition is observed and is of spin-flip type. In this case, the adsorbate spin is initially coupled to the $B$ field, it couples to the tunnelling electron spin during the collision and couples back to the $B$ field at the end. On this example, one thus sees the difference between the non-spin-flip 0-1 transition at low $B$ which implies the transient decoupling of the local spin from the substrate-induced anisotropy and the spin-flip 0-2 transition at large $B$ which implies the transient decoupling of the local spin from the magnetic field. The gradual change of excitation probabilities seen in Fig.~\ref{figure_Fe-Th} thus corresponds to the gradual switch from a structure dominated by  the magnetic anisotropy of the adsorbate to a structure dominated by  the applied $B$ field. In this way, the good agreement between theory and experiment on the inelastic intensities is a further proof of the validity of the description of the Fe magnetic structure by a local spin and the anisotropy Hamiltonian (Eq.~\ref{anisotropy})

\subsubsection{Link between the various approaches}

In the theoretical approaches discussed above, it could seem that
the final results for inelastic transitions were different; indeed,
in one case ($S^2$ theories), the inelastic transitions are found to be
proportional to squared matrix elements of $\vec{S}$, the adsorbate spin,
whereas in the other (strong coupling approach), everything is expressed
via the total spin, $\vec{S}_T$, the total spin of the system (electron
+ adsorbate ). In fact it is possible to show that
the predictions of Persson's approach~\cite{Mats_PRL2009} are equivalent
to those of the strong coupling approach, though without a full account
of the system symmetries. In Persson's approach, the 'anisotropy-free'
amplitude for tunnelling through the adsorbate is expressed as

\begin{equation}
T = T_0  + T_1  \vec{s} \cdot \vec{S}
\label{TMats}
\end{equation}
with two parameters, $T_0$  and $T_1$. We can reexpress the above
tunnelling amplitude by introducing the total spin, $\vec{S}_T = \vec{S}
+\vec{s}$ instead of the scalar product, ($\vec{s} \cdot \vec{S}$):

\begin{equation}
T = T_0 + 0.5 \, T_1 \times ( \vec{S}_T^2 - \vec{S}^2 -\vec{s}^{\; 2}).
\label{TMats2}
\end{equation}

If we further introduce a closure relation on the eigenstates of $\vec{S}_T^2$ and ${S}_{Tz}$ (quantum numbers $S_T$ and $M_T$), we get:

\begin{eqnarray}
T  &=&   \sum_{M_T} \left|S_T=S+1/2, \, M_T\right\rangle 
(T_0  +  0.5\, T_1 \,  S) 
\nonumber \\
&\times& \left\langle S_T=S+1/2, \, M_T\right|   + 
\sum_{M_T} \left|S_T=S-1/2, \, M_T\right\rangle \nonumber \\ 
&\times& 
(T_0  - 0.5 \, T_1 \,  (1 + S)) \left\langle S_T=S-1/2,M_T\right|  
\label{Tequiv}
\end{eqnarray}
where $S(S+1)$ is the eigenvalue of $\vec{S}^2$. Equation~(\ref{Tequiv})  is exactly of the same form as Eq.~(\ref{Tmatrix}), expressing the tunnelling amplitude as an expansion over the different ($S_T$, $M_T$) symmetries:
\begin{eqnarray}
T &=& \sum_{M_T} \left|S_T=S+1/2,M_T\right\rangle 
T^{S+1/2} \left\langle S_T=S+1/2,M_T\right| \nonumber \\  &+& 
\sum_{M_T} \left|S_T=S-1/2,M_T\right\rangle 
T^{S-1/2} \left\langle S_T=S-1/2,M_T\right|  
\label{Tequiv2}
\end{eqnarray}
with the connection formulae: 

\begin{eqnarray}
\label{Equiv2}
T^{S+1/2} &=&
T_0  +  0.5 \: T_1 \: S\nonumber \\ 
 T^{S-1/2} &=& T_0  - 0.5 \: T_1 \: (1 + S) 
\label{Lien}
\end{eqnarray}

Equation~(\ref{Lien}) then provides the formal link between
Persson's approach~\cite{Mats_PRL2009} and the strong coupling
approach~\cite{Lorente_PRL2009,Novaes_FePc}. All factors, $T_0$,
$T_1$, $T^{S+1/2}$ and $T^{S-1/2}$ are  complex numbers and we can
switch from one representation to the other. It also allows to make
the link between the strong coupling approach and the extended $S^2$
theory (\cite{Loth_NJP,Chilian_2011}) using an adjusted ($u  +  {\vec{s}} \cdot {\vec{S}}$)
transition operator. In that case, the adjusted $u$ term only contributes
to elastic tunnelling  and the above equivalence thus explains the success
of  the original $S^2$ approach in describing the relative population
of excited channels.

At this point, one can stress the differences for practical use of
the equivalent equations ~(\ref{TMats}) and ~(\ref{Tequiv2}). In the
cases we studied (FePc, CoPc, Fe and Mn adsorbates~(\cite{Lorente_PRL2009,Novaes_FePc}),
tunnelling through the adsorbate is dominated by a single $S_T$ symmetry,
so that a single term has to be be used in equation~(\ref{Tequiv2}). Actually, such a situation is the one
 in which tunnelling is dominated by a single spin-orbital of the system. 
The $|T^{S_T}|^2$ term appears as a global factor in the conductance
and the relative intensities in the various elastic and inelastic channels
can be obtained simply from spin-coupling coefficients. In contrast,
for the same situation, using  Eq.~(\ref{TMats}),  both terms
in~(\ref{TMats})   contribute to tunnelling and only an ab initio
detailed study of tunnelling can predict the relative weight of the
two. As a further example, defining the $S_T$ tunnelling channel yields
the values of $u$ in the extended $S^2$ formulation that would correspond
to tunneling via a single $S_T$ value: $u$ = ($S$ +1)/2 for  $S_T$ = $S$
+ 1/2 and $u$ = - $S$ for $S_T$ = $S$ - 1/2. This clearly illustrates the
advantage of fully expressing the symmetry of the tunnelling process,
i.e. of introducing the $S_T$ tunnelling channels, as performed by the
strong coupling approach.

\subsection{Tunnelling symmetry and Single Atom Magnetisation Curve (SAMC)}

As discussed above, the  tunnelling symmetry (value of the total spin $S_T$) is extremely important for determining the magnitude of the inelastic magnetic transitions. However, it also influences other experimentally observable quantities implying polarised electrons. Indeed, measurements of the tunnelling current above an adsorbed atom with a polarised tip yields direct information about the magnetic polarisation of the adsorbate (see e.g. 
Refs.~\cite{KhajetooriansPRL2011,MeierNature2008,YayonPRL2007,IacovitaPRL2008,KhajetooriansScience2011,ZhouPRB2010,SerrateNature2010} and discussions therein).
Among these studies, measurements of Single Atom Magnetisation Curve (SAMC) present some appealling advantages. They consist in  measuring the tunnelling current with a polarised tip above the adsorbed atom as a function of an applied magnetic field~\cite{MeierNature2008}. For this, the tip polarisation has to survive to the applied magnetic field, i.e. to remain the same over the whole range of studied B field. If one uses a classical approach for the field dependence of the current, assuming that the tunnelling current contains a term proportionnal to the scalar product of the tip magnetisation and of the  adsorbate spin, then SAMC measurements provide  direct information on the polarisation of the adsorbate, its direction and the associated spin value. Actually, a SAMC presents the transition between the current for parallel and anti-parallel polarisations of the tip and adsorbates. The actual shape of the transition depends on the system temperature as well as on the magnetic anisotropy of the adsorbate. Analysis of SAMC curves imply the fit of the transition curve using a classical approach.

\begin{figure}
\centering
\includegraphics[angle=270,width=0.6\textwidth]{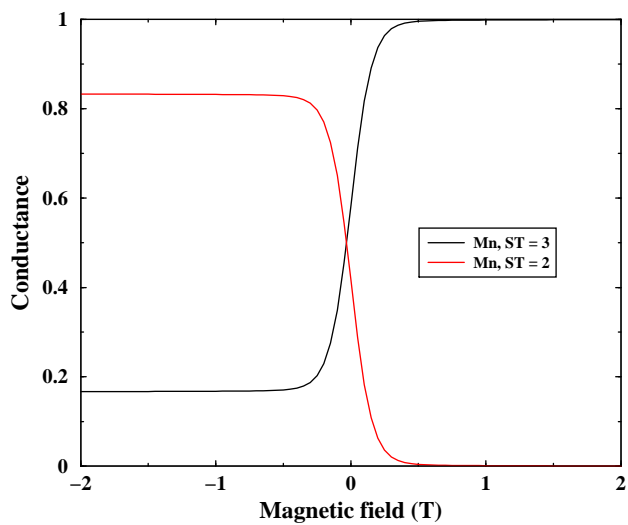}
\caption{\label{figure_SAMC}
Single Atom Magnetisation Curve (SAMC) for a single Mn adsorbate on a CuN/Cu(100)surface. The current is computed in  the strong coupling approach,  using the earlier modelling of the Mn adsorbate anisotropy and spin value (Mn adsorbate spin $S$ = 2.5)~\cite{HeinrichScience2007,Lorente_PRL2009}. The tip is assumed to be fully polarised, the junction bias is larger than all the inelastic magnetic thresholds  and the surface temperature is 0.5 K. The SAMC is computed for the two possible values of the spin symmetry of the tunnelling process: $S_T$ = 2 and 3.  } 
\end{figure}

However, the discussion presented above about the effect of the spin symmetry ($S_T$ symmetry of the tunnelling process) on the inelastic magnetic transitions can be transposed to the calculation of SAMC. Indeed, if we know the degree of polarisation of the tip (e.g. the fraction of electrons with up and down spins), the strong coupling approach can yield the tunnelling current as a function of the applied B field. Figure~\ref{figure_SAMC} presents the SAMC for a single Mn atom on a CuN/Cu(100) surface, computed using the strong coupling modelling used for the treatment of the magnetic excitation in this system~\cite{Lorente_PRL2009}. The calculation of the current dependence on the B field is thus fully quantal. The tip is assumed to be fully polarised parallel to the direction of the B field, the junction bias is assumed to be larger than  all the inelastic magnetic thresholds and the surface temperature is equal to 0.5 K.

Figure~\ref{figure_SAMC} presents the results of the strong coupling calculations for the two possible values of the tunnelling spin symmetry, $S_T$ = 2 and 3 (the actual value for the Mn/CuN/Cu system is $S_T$=3, see Ref.~\cite{Lorente_PRL2009}). The two SAMC appear to be very different, exhibiting opposite behaviours when the applied B field is varied. This illustrates the effectiveness of the $S_T$ symmetry in the tunnelling process. Incident electrons with spin up and down  are differently favored in the two $S_T$ symmetries, leading to the reversal of the SAMC shape.  
It shows that the spin dependence of the electron transmission through a magnetic adsorbate is more complex than a mere discussion  in terms of  majority or minority spins, the details of the spin coupling between tunnelling electron and adsorbate have actually to be considered. Finally this also shows that the
  key parameter for describing the strength of magnetic inelastic transitions, the $S_T$ symmetry, could be extracted from a spin-polarised SAMC, which does not resolve the various inelastic transitions.

\subsection{Comparison with vibrational IETS: enhanced
efficiency in magnetic IETS}

Magnetic IETS experiments, as well as the strong coupling approach
results, showed that magnetic transitions are extremely easily induced by
tunnelling electrons. In several cases, inelastic tunnelling was found to
dominate over elastic tunnelling, a feature at variance with vibrational
IETS, where inelastic tunnelling represents at most a few per cent of
elastic tunnelling~\cite{StipeScience1998,LorentePRL2000,Komeda,Ho2002}. This feature can
be understood in the framework of the strong coupling approach. Indeed,
as seen in Eq.~(\ref{P3}), the relative populations of the final
states depends on the relative value of the $A^*_{j,n',m'}$ coefficients,
i.e. on the relative weights of the final states in the intermediate $j$
state. As discussed above, the active $S_T$ symmetry (the $j$ state)
acts as a filter for the tunnelling and the relative weights of the
final states depend on their weight in the intermediate state, i.e. in
the tunnelling $S_T$ symmetry. In this way, the process obeys  spin
selection rules and allow conductance 
cases dominated
by inelastic channels.

Actually, the process as formulated in the strong coupling approach
bears strong resemblances with other processes involving an angular
momentum exchange in atomic, molecular or surface science. As a first
example, the resonant rotational excitation of $H{_2}$ molecules by low
energy electron impact has been formulated many years ago by Abram et
Herzenberg~\cite{Herzenberg} in a way very close to the magnetic
IETS strong coupling approach. In low energy electron collisions
on molecules, scattering is strongly influenced by a few resonances
corresponding to the transient capture of the collisional electron
by the target molecule. These are known to dominate the vibrational
excitation process at low energy~\cite{Schulz}. The resonances are
also associated with a well defined molecular symmetry, i.e. with a
projection of the electron angular momentum on the intermolecular axis;
very often they almost correspond to  a single angular momentum in the
molecular frame, $p \sigma$ wave in the case of low energy electrons
colliding on an $H{_2}$ molecule discussed in Ref.~\cite{Herzenberg}. The dominance of a single angular momentum in resonant scattering parallels the dominance of a single $S_T$ total spin in the strong coupling description of magnetic IETS. In
the resonant scattering process, the incident electron then brings a well
defined angular momentum which couples to the molecule angular momenta;
at the end of the collision, the resonance decays by emitting an electron
with a well-defined angular momentum. This coupling/decoupling sequence
between the electron and molecule angular momenta leads to an efficient
angular momentum transfer between the collision partners. This process
can be easily expressed in the sudden approximation since the rotational
motion of the molecule is much slower than the electron collision, even
resonant.  This sudden approximation approach for rotational excitation
has been much used in the field of electron-molecule collisions and the
momentum exchange process outlined above was found to account very often
for the observed excitation. This process also exists in the case of
adsorbed molecules and lead to an efficient excitation of the frustrated
rotational motion motion as discussed in ~\cite{BillyO2,Ho_O2}. The
above examples concern rotational excitation, i.e. the transfer of orbital
angular momentum. However, a similar process for spin angular momentum
exchange (like in magnetic IETS) has also been invoked in the case of
excitation of forbidden transitions in electron-molecule transitions,
both in gas phase collisions~\cite{Billy} and in the case of scattering
on surfaces~\cite{Bahrim1, Bahrim2}.
\textcolor{red}{Finally, one can also mention an experimental study of transitions between spin-orbit components in low energy electron collisions on NO molecules by M.Allan~\cite{AllanPRL2004,Allan2005}. The ground state of NO is a $^2\Pi$ electronic state which splits into two spin-orbit components: $^2\Pi_{1/2}$ and $^2\Pi_{3/2}$. At low energy, a $^3\Sigma$ resonance dominates e-NO scattering. Resonant scattering in the $^3\Sigma$ symmetry thus corresponds to a decoupling /recoupling sequence between the molecular spin and the collisional electron spin. It closely parallels the magnetic excitation process as viewed in the strong coupling approach and as a result, the $^3\Sigma$ resonant scattering is inducing strong inelastic transitions between the two spin-orbit components of the NO ground state~\cite{AllanPRL2004,Allan2005}.}

\textcolor{red}{The angular momentum exchange process with a collisional electron } thus seems to be of common
occurrence in free molecule and surface problems and to be always
of high efficiency. One can stress that the formulation used in the
strong coupling approach of the magnetic IETS closely parallels the
treatment of the rotational excitation in electron-molecule collisions  in
~\cite{Herzenberg}; in both cases, it is based on a sudden approximation
associated to the explicit account of the angular momentum symmetry of
the intermediate state and this allows to obtain the relative elastic
and inelastic populations as ratios of spin coupling coefficients. In
a way, it can be described as a recoil phenomenon; in the Abram and
Herzenberg case, the electron in the initial and final states is associated with
a well-defined angular momentum and via angular momentum conservation,
scattering has to be associated to an 'angular recoil' of the molecule
i.e. to rotational excitation.  One can then wonder about the possibility
of a similar process for vibrational excitation, i.e. for the exchange
of linear momentum. In  that case a recoil momentum transfer exists. Let
us consider an electron scattered by a free molecule, the direction
of the velocity of the electron changes in the scattering and thus the electron
transfers some linear momentum to the molecule. This linear momentum recoil  is the equivalent
of the process discussed above for angular momentum recoil. However, due to the very large mass ratio
between electrons and nuclei, the energy that can be actually transfered
by a low energy electron is very weak and cannot excite   vibrational
motion. The situation is completely different for angular momentum
exchange; due to quantization, electron and molecule angular momenta are
of the same order of magnitude, allowing efficient exchange processes.
Vibrational excitation then invokes usually another excitation process:
resonant vibrational excitation~\cite{Schulz,Gadzuk1983,Palmer_1992,Djamo_1993}; in that case, the
long lifetime of the resonance allows a significant energy transfer
between electron and nuclei, in a way during the resonance lifetime many
electron-molecule interactions can occur. We can then conclude that the
very efficient mechanism at play in magnetic excitations, also exists
for vibrational excitations but that it is not efficient.

\subsection{Excitation of magnetic adsorbates by polarized electrons, spin-transfer torque}

The works described above show that the probability of magnetic excitation
of an individual adsorbate by a tunnelling electron can be very  large. If
the tunnelling electrons are spin-polarized, then one can expect that
tunnelling electrons are able to transfer  part of their polarisation to
the adsorbate. Indeed, electron transmission through the adsorbate can
be a spin-flip or non spin-flip process; the latter does not change the
adsorbate polarisation whereas the former tends to align the adsorbate
polarisation with that of the tunnelling electron, i.e. to populate some
specific excited states; this spin transfer process is counterbalanced
by the decay of the magnetic excited states which bring the system back
to its ground state.  This phenomenon is usually called 'spin-transfer
torque' by reference to the capability of tunnelling electrons to rotate
the adsorbate polarisation.

\begin{figure}
\centering
\includegraphics[width=0.4\textwidth]{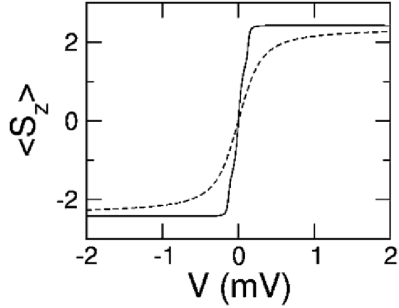}
\caption{\label{fig2a_prl_delgado_2010}
Model calculation of the magnetisation of a Mn adsorbate on a CuN/Cu(100) substrate. For each bias value, the steady state magnetisation of the adsorbate (mean value of $\left\langle S_z \right\rangle$) is evaluated by solving equation (\ref{eq_master}) for two temperatures:  
$T= 1$ K (dashed line) and $T=0.1$ K (solid line). The tip magnetisation is parallel to the Mn easy axis. Reprinted with permission from \cite{DelgadoPRL2010}. Copyright 2010 by the American Physical Society.
}
\end{figure}

The spin-pumping effect has been described using a rate equation
formalism~\cite{LothNatPhys2010,DelgadoPRL2010,NovaesPRB2010}: the population of the various states of the adsorbates evolves
with time due to excitation and de-excitation processes induced by the
tunnelling current and by the spontaneous  relaxation of the magnetic
excitations.  For an STM tip positioned above the  adsorbate, a tip
bias V and a tunneling current I, the time dependence of the population,
$P_i(I,V)$, of the magnetic state, i, is given by:

\begin{eqnarray}
 \frac{dP_i(I,V)}{dt}=&-&P_i(I,V)\left(\sum_j \Gamma_{i,j} + \sum_j F_{i,j}(I,V) \right) \nonumber \\
                      &+& \sum_j P_j(I,V) \left( F_{j,i}(I,V)+\Gamma_{j,i}\right),
\label{eq_master}
\end{eqnarray}
where $\Gamma_{i,j}$ is the partial decay rate of state i towards state j.
 $F_{i,j}(I,V)$  is
 the transition rate from state i to state j induced by the tunneling
electrons.

The populations given by equation~(\ref{eq_master}) very quickly reach
a steady-state equilibrium, balance between spin-pumping and relaxation
effects. The typical time for reaching this steady state is given by
the order of magnitude of the relaxation time; as discussed in the next
section, it is in the sub-ns range for the Mn/CuN/Cu(100) system going
down to the fs range for magnetic atoms adsorbed directly on a metal. This
is much faster than the typical time scale of an STM experiment which
can then be described by the steady-state equilibrium (note, however,
the existence of ultrafast STM experiments which can study magnetism
dynamics in real time~\cite{LothScience2010}). The first theoretical
study for individual adsorbates using equation~(\ref{eq_master}) was
reported by Delgado et al~\cite{DelgadoPRL2010, Delgado_PRB2010} on a
model case; then it was further applied to the case of Mn on CuN/Cu(100)
in connection with an experimental study of spin relaxation times via
saturation effects~\cite{LothNatPhys2010,NovaesPRB2010}. The spin torque
effect can be very effective. Figure~\ref{fig2a_prl_delgado_2010}
from \cite{DelgadoPRL2010} presents the steady-state
magnetisation of a Mn adsorbate (spin $S$ = 5/2) as a function of the
applied tip bias. It was computed using a set of parameters describing
pumping and relaxation for a perfectly polarized tip. It appears
that the magnetisation of the adsorbate can be easily reversed by
the tunnelling current, i.e. by  purely electrical means, yielding a
control of the adsorbate magnetisation. In the case shown in Fig.~\ref{fig2a_prl_delgado_2010} with a
perfectly polarized tip, the saturation polarization at 'large' bias is
almost complete. This complete switch of the adsorbate magnetisation
is associated with a significant change of the adsorbate conductance,
which can then yield the experimental signature of the pumping effect.

\begin{figure}
\centering
\includegraphics[width=0.6\textwidth]{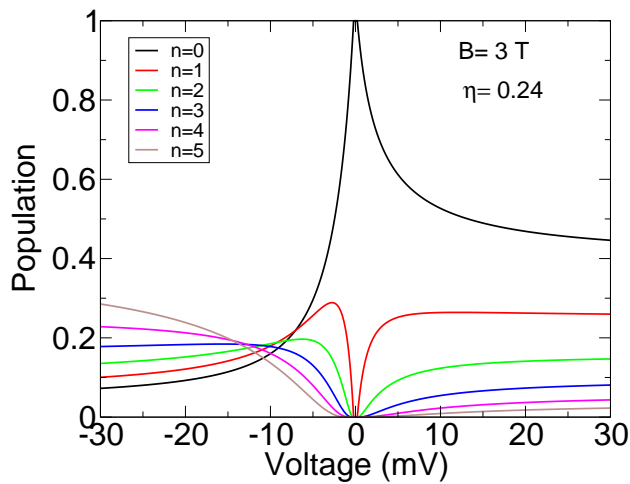}
\caption{\label{populations_pol}
Population of the six magnetic states in the Mn/CuN/Cu(100) system as a
function of the tip bias. The magnetic field is along the x-axis and equal
to 3T. The tip polarisation, $\eta$, is equal to 0.24. The conductance
at zero bias is equal to $2. 10^{-6}$ S. \textcolor{red}{ Taken from reference~\cite{NovaesPRB2010}.}
}
\end{figure}

\begin{figure}
\centering
\includegraphics[width=0.6\textwidth]{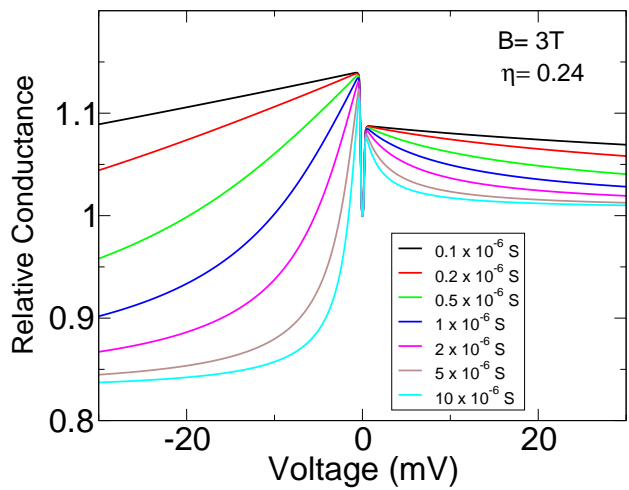}
\caption{\label{cond_3T}
Relative conductance of the Mn/CuN/Cu(100) system as a function the tip
bias. The tip polarization is $\eta = 0.24$ and the B field, equal to
3T, is along the x-axis. The finite population of the excited states is
taken into account. The various curves correspond to various absolute
conductances at zero bias (0.1, 0.2, 0.5, 1., 2., 5. and 10. $10^{-6}$
S). In the figure, the conductance is plotted in relative value, with the
conductance for zero bias set to 1. \textcolor{red}{ Taken from reference~\cite{NovaesPRB2010}.}
}
\end{figure}

The possibility to generate a steady state population of excited
states in an adsorbate allows some control on the adsorbate by
electrical currents. This phenomenon has been used by Loth et
al~\cite{LothNatPhys2010} to determine the relaxation times in the
case of Mn adsorbates on CuN/Cu(100) (section~\ref{experiments}). The idea
is that a significant current creates a steady state population of
excited states that modifies the adsorbate conductance, a careful
modelling via equation~(\ref{eq_master}) then allows the determination
of the $\Gamma_{i,j}$ relaxation times~\cite{LothNatPhys2010}. The
steady-state population of the Mn states have been modelled
in Ref.~\cite{NovaesPRB2010} using ab initio computed rates for
equation~(\ref{eq_master}). Figure~\ref{populations_pol} presents the population of the six states of Mn
(0 is the ground state and 5 the highest excited state, polarised
almost anti-parallel to the ground state)) as a function of the tip
bias for conditions typical of experiment~\cite{LothNatPhys2010}
(a partially polarised tip, magnetic field of 3 T). Consistently with
Fig.~\ref{populations_pol}, there exists a significant population of excited
states induced by the tunnelling current that is asymmetric with respect
to the tip bias.  Note that for the considered currents (conductance of
2.0 $10^{-6}$ S), the population of magnetic states is drastically changed
by the electron pumping effect, excited state populations dominating over
the ground state one at large bias. This leads to drastic changes in the
polarized electron conductance of the adsorbate that have been observed
experimentally and theoretically~\cite{LothNatPhys2010,NovaesPRB2010,
Delgado_PRB2010}. Figure~\ref{cond_3T} presents the
conductance of an individual Mn adsorbate on CuN/Cu(100) as a function
of the junction bias for several values of the conductance at zero bias
(this corresponds to different tip-adsorbate distances as well as to
different currents flowing through the adsorbate). In this system,
the magnetic thresholds are responsible for the structure at very small
bias below 1 meV, whereas the variation of the conductance as the bias
is further increased is due to the increase of excited state populations
(see Figure~\ref{cond_3T}). This variation is a direct consequence of
the balance between current-dependent excitation and deexcitation induced
by the electrons and current-independent spontaneous relaxation of the
magnetic excitation. The use of  polarized electrons  allows to reveal
the existence of a steady state population of excited polarized states,
i.e. of the spin-transfer torque effect discussed above.

In the case of non-polarized tunnelling electrons,
a large population of excited state is also generated,
however symmetrically for positive and negative biases (see
e.g. in~\cite{DelgadoPRL2010,NovaesPRB2010,Loth_NJP}). In addition,
in this case, the adsorbate conductance for large biases, far above
the inelastic magnetic thresholds, is independent of the existence of
a significant excited state population, as discussed theoretically and
observed experimentally~\cite{LothNatPhys2010,NovaesPRB2010,Loth_NJP}. This
feature can be linked with the qualitative view of the excitation process
as a sharing among the possible final states\cite{NovaesPRB2010} and
is at variance with the large variations of conductance observed in the
polarized electron case (see  Fig.~\ref{cond_3T})).

\subsection{Evaluation of magnetic state lifetimes: decay via electron-hole
pair excitation}

All the experimental results and theoretical studies discussed above point
at the extreme efficiency of a tunnelling electron in inducing magnetic
transitions in a local spin carried by an adsorbate. Actually, as is
clearly seen in the strong coupling approach, the tunnelling character
of the electron is not essential, in fact any electron colliding with a
magnetic atom can be very efficient in inducing magnetic transitions. As
a consequence, there should exist a very efficient quenching mechanism
for an excited local spin on a surface: electrons coming from the
substrate continuously hit the adsorbate and are scattered back into
the substrate; these electrons \textcolor{red}{can a priori be} very efficient in inducing
magnetic transitions and thus lead to an efficient quenching mechanism
via electron-hole pair creation (an electron from the substrate with an
energy lower than Fermi energy is scattered super-elastically from the
excited adsorbate and goes back into the substrate with an energy above
Fermi energy). This mechanism has been invoked for the interpretation
of the lifetime measurements in the case of local spins on individual
adsorbates~\cite{LothNatPhys2010,LothScience2010}.

An extension of the strong coupling approach  has been presented
for the treatment of magnetic state decay via electron-hole pair
creation~\cite{NovaesPRB2010}. It directly elaborates on the treatment
outlined in section~(\ref{Strong}) for magnetic excitation. It makes
use of the same anisotropy Hamiltonian (\ref{anisotropy}) and of its
eigenstates $\phi_n$ associated to the $E_n$ eigenenergies. The decay
rate, $\Gamma_{Tot,i}$, of an excited state, $|\phi_i\rangle$, with
energy $E_i$, is the inverse of its lifetime $\tau_i$ and it can then
be written using matrix elements of the T transition matrix (we assume
the energy variation of the T matrix to be small on the energy scale of
the $i \rightarrow f$ transition and we assume a vanishing temperature
of the substrate)~\cite{Lorente08},

\begin{eqnarray}
 \frac{1}{\tau_i} 
 &=& \Gamma_{Tot,i} = \sum_f \Gamma_{i,f} = \sum_f \frac{2\pi\delta\Omega_f}{\hbar} \nonumber \\ 
&\times& \sum_{\substack{k_i,k_f \\ m_i,m_f}} \left|\langle 
k_f,m_f,\phi_f |\hat{T}| k_i,m_i,\phi_i \rangle\right|^2  \nonumber \\
&\times& \delta(\varepsilon_i-\varepsilon_f) \delta(\varepsilon_i-E_F),
\end{eqnarray}

where $|\phi_f\rangle$ are the final states of the decay, associated to
an energy transfer of $\delta\Omega_f=E_i-E_f$, and the total energy is
$E_T=E_i+\varepsilon_i=E_f+\varepsilon_f$. The initial and final states of
the substrate electrons are noted by their wave numbers, $k_i$ and $k_f$,
and by their initial and final spin projections on the quantization axis,
$m_i$ and $m_f$. Each term, $\Gamma_{i,f}$, in the sum over $f$ is
the partial decay rate of the initial state to a peculiar final state.
The transition T-matrix can be expressed using the sudden approximation
using the T-matrix defined in the absence of magnetic anisotropy (in
the absence of spin-orbit coupling). Similarly to the excitation case
(see section~\ref{Strong}), the latter is then expressed as a sum
over several terms corresponding to the different tunnelling symmetries,
i.e. to different $S_T$ and $M_T$ to yield:

\begin{eqnarray}
 &\frac{1}{\tau_i} 
&= \, \sum_f \frac{2\pi \, \delta\Omega_f}{\hbar} \sum_{\substack{k_i,k_f \\ m_i,m_f}} \delta(\varepsilon_i-\varepsilon_f) \delta(\varepsilon_i-E_F)   \\ 
&\times& \left|\sum_{S_T} \langle k_f|T^{S_T}|k_i\rangle \sum_{M_T}\langle m_f,\phi_f|S_T,M_T\rangle \langle S_T,M_T|m_i,\phi_i \rangle\right|^2. \nonumber
\end{eqnarray}

In the above, all the possible values of the total spin, $S_T$, are
included. If we assume that only one spin symmetry is contributing
effectively to the magnetic quenching, the expression can be further
simplified to yield~\cite{NovaesPRB2010}:

\begin{eqnarray}
 \frac{1}{\tau_i} =  \sum_f \Gamma_{i,f} = \sum_f \frac{\delta\Omega_f}{h} 
  (2\pi)^2 T^{S_T}(E_F) P_{Spin}(S_T,i\rightarrow f).
  \label{Quenching}
\end{eqnarray}
A simple expression similar to that found for excitation
(equation~(\ref{P3})) is then recovered. In the case where the two $S_T$
symmetries contribute to the quenching, one can make a statistical
approximation justified by the summation over many substrate states and
still use an expression similar to ~(\ref{Quenching}):

\begin{eqnarray}
 \frac{1}{\tau_i} =  \sum_f \Gamma_{i,f} = \sum_f \frac{\delta\Omega_f}{h} 
  (2\pi)^2 T(E_F) P_{Spin}(i\rightarrow f).
  \label{Quenching2}
\end{eqnarray}
the term $T(E_F)$ then corresponds to the total (all $S_T$ symmetries)
flux of substrate electrons, and is given by:
\begin{eqnarray}
T(E_F) = (2\pi)^2 \sum_{k_i, k_f} |\langle k_i|T| k_f \rangle|^2 \delta (\varepsilon_i-\varepsilon_f) \delta ( \varepsilon_i - EF),
\label{Te}
\end{eqnarray}
which can be easily implemented in DFT~\cite{NovaesPRB2010}.
The probability $P_{Spin}(i\rightarrow f)$ is
defined as an average over the two $S_T$ values.

The physical meaning of equations~(\ref{Quenching})
and~(\ref{Quenching2}) is transparent, it can be interpreted as the
product of the number of electrons hitting the adsorbate per second in
the appropriate energy interval by a spin-transition probability. The
spin-transition probability is expected to be significant and the
flux factor, $T(E_F)$, to depend on the system under investigation. In
particular, one can expect the presence of an insulating ultra-thin layer
between the adsorbate and the substrate to reduce the flux factor.

The flux factor bears strong resemblances with the tunnelling
term in the excitation formalism. Actually, it has been
shown~\cite{NovaesPRB2010} that it can be calculated from first
principles using an approach very similar to the one used in the
TRANSIESTA code to compute tunnelling fluxes. The idea is to introduce a
single reservoir (the substrate) instead of two (the substrate and the
tip) and to compute the flux from the reservoir into the reservoir via
the magnetic atom; the latter is defined by a set of atomic orbitals.

\begin{figure}
\centering
\includegraphics[width=0.6\textwidth]{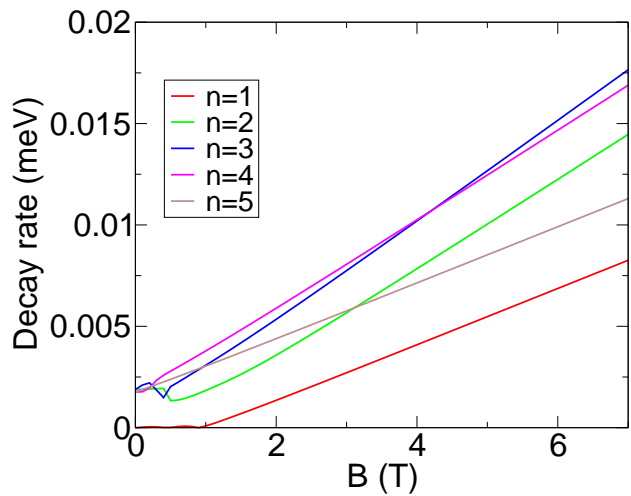}
\caption{\label{fig5}
Decay rate (in meV) of the five
excited magnetic states of a Mn atom on CuN/Cu(100) as a function
of an external magnetic field $B$. \textcolor{red}{Taken from reference~\cite{NovaesPRB2010}.} 
}
\end{figure}

\begin{figure}
\centering
\includegraphics[width=0.6\textwidth]{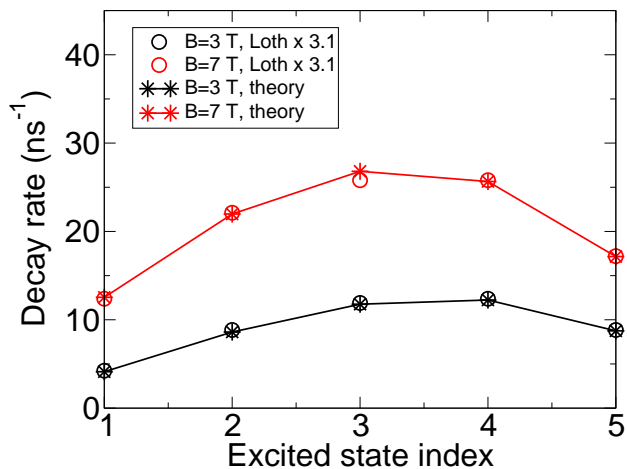}
\caption{\label{decay_rates_comp}
Comparison between the \textcolor{red}{theoretical decay rate (stars)} of the five excited
magnetic states in the Mn/CuN/Cu(100) system with the experimental
results of Loth et al~\cite{LothNatPhys2010} \textcolor{red}{(open circles)}. The black symbols 
 correspond to an applied B-field of 7T and the red symbols 
to 3T (the B-field is along the x-axis).  \textcolor{red}{The lines are only guides for the eye. }For the sake of comparison,
the experimental results of Loth et al have been multiplied by a global
factor equal to 3.1. \textcolor{red}{Taken from reference~\cite{NovaesPRB2010}.}
}
\end{figure}

 This approach has been used in the case of Mn adsorbates on CuN/Cu(100)
 studied experimentally by Loth et al~\cite{LothNatPhys2010}. The
 corresponding results for the decay rate as a function of the applied B
 field are shown in Fig.~\ref{fig5}. The spin, $S$,
 of the Mn adsorbate is 5/2, so that there are five excited states. As
 seen in Fig.~\ref{decay_rates_comp}, the decay rate of the lowest lying state is very small
 at small B; this is a direct consequence of the adsorbed Mn magnetic
 structure and not so much a sign of stability of the state: the lowest
 state has a very small excitation energy at low B, so that its decay
 rate is very small (see Eq. ~(\ref{Quenching})). Besides this
 energy effect, typically, the decay rates are in the range of 2 to 20
 $\mu$eV, corresponding to lifetimes of the order of 0.3-0.03 ns. One can
 notice that the decay rates increase with the applied magnetic field,
 almost linearly at large B. This is also an  effect of the de-excitation
 energy; in the Zeeman limit at large B, the energy difference between
 the states vary linearly with B and so do the decay rates.

 These magnetic lifetimes are rather short, so that it seems
   likely that the electron-hole pair creation process will dominate the
   excited state decay over other processes, involving e.g. atomic
   motions~\cite{FabianSarma1999}. Figure~\ref{decay_rates_comp} presents a comparison between the strong coupling results and
   the experimental results~\cite{LothNatPhys2010,NovaesPRB2010}. The
   relative values of the decay rates (ratio between excited states,
   dependence on B) is well accounted for, but the theoretical results
   are a factor 3.1 larger than the experimental data. The origin of
   this discrepancy is not clear.

 The theoretical estimate for the decay rate
 (expression~(\ref{Quenching2})) can be used to discuss the possible
 variations of the magnetic state lifetime, i.e. of the relaxation time
 of a local magnetic moment on an adsorbate. Besides the excitation
 energy term, already discussed, two effects could lead to very different
 lifetimes, short or long. First, the flux factor can be very different
 on different substrates; for example, a 'thick' layer of an insulator
 could effectively decouple the surface from the underlying metallic substrate
 and  significantly decrease the flux at the adsorbate leading
 to very long-lived states. Similar effects have been invoked to
 explain long-lived states (non-magnetic, delocalised or localised) at \textcolor{red}{metal} surfaces in the presence
 of insulating layers~\cite{Harris,Wolf,Marinica,Wurth2000,Gauyacq2004}. Another variation could
 come from the spin-probability factor. Certain spin structures could be
 designed such that transitions induced by electrons between  an excited
 state and the ground state are \textcolor{red}{forbidden or }quasi-forbidden; this can happen e.g. in
 the case of a local spin with a large anisotropy (large negative D in
 Hamiltonian~(\ref{anisotropy})) at finite B field. In this case, the
 two lowest lying  states are quasi-Zeeman states (almost eigenstates of
 $M_z$), which do not fulfill the ${\Delta}M$ = 0,$\pm$1 selection rule. This
 structure has been suggested in Ref.~\cite{HeinrichScience2007,Freedman}  as a possible model
 for bistable systems; the  long lifetimes found by ~\cite{LothScience2010}
 in the case of Fe and Cu adsorbed on CuN/Cu(100)
 also pertains to this class of systems.

In this context, one can mention a series of theoretical studies
devoted to individual 3d magnetic atoms adsorbed directly on a metal
substrate~\cite{ KhajetooriansPRL2011,Lounis2003,Lounis_PRL2010,Lounis_PRB2011}. In
that case, the coupling between adsorbate and substrate is much stronger
than in the case discussed above that involved a partly insulating
layer between adsorbate and metal. This leads to a strong hybridization
between the atom and the metal and ultimately to much reduced lifetimes
for the magnetic excitations. Theoretically, these systems were treated
using an itinerant magnetism kind of approach instead of the local spin
concept discussed above. The susceptibility of the adsorbate/substrate
system, response to an applied oscillating transverse magnetic field,
was computed taking into account the effective interaction between electrons. The 
susceptibility as a function of the excitation frequency displays
peaks at the allowed excitation energies, the width of which corresponds
to the excited state lifetimes. The decay of the excited magnetic
states is again due to electron-hole pair creation, called Stoner
excitations in this context. In the case of Fe adsorbates on Cu(111),
also studied experimentally by magnetic IETS~\cite{KhajetooriansPRL2011},
the theoretical lifetime of the excited magnetic state was found to be
very short, typically in the 500 fs range, in excellent agreement with
the measured linewidth. The level width is seen to vary almost linearly
with the applied magnetic field, basically following the variation of
the  excitation energy of the state (see above a discussion  of this
effect for local spins). Actually, it appears that the level width of the
excited magnetic  state is almost equal to the excitation energy of the
state, confirming its highly unstable character and explaining why the
excited state is only  observed as a very broad structure and not as a
sharp step in the conductance in a magnetic IETS experiment.

\subsection{Chains of magnetic atoms and spin wave excitation}
\begin{figure}
\centering
\includegraphics[width=0.6\textwidth]{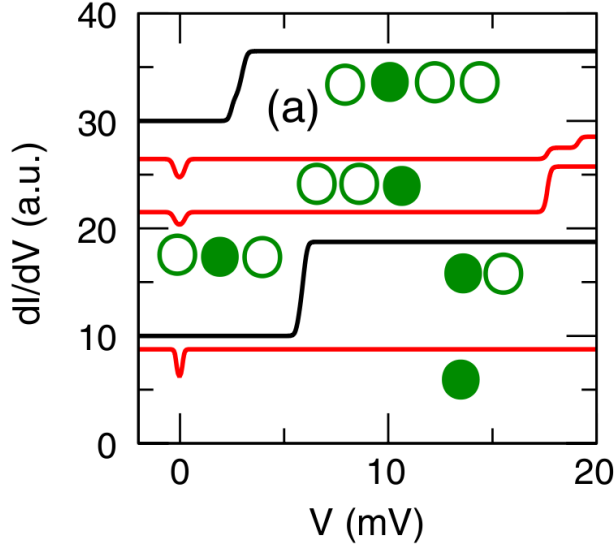}
\caption{\label{fig2a_prl_rossier_2010}
$dI/dV$ for $B=0$, $ k_BT=0.6$ K for
chains of N Mn atoms, $N=1, 2, 3, 4$. Solid circles represent
the atom underneath the tip. Reprinted with permission from
\cite{FernandezRossier_PRL2009}. Copyright 2009 by the American Physical
Society.
}
\end{figure}

All the theoreoretical studies discussed above considered the case of a single magnetic adsorbate on a surface, carrying a local spin. In connection with experiments on chains of magnetic adsorbates, several theoretical studies addressed the problem of chain excitation and its links with spin-wave excitation in an infinite 1D-spin chain.
The first study considered the case of short linear chains of Mn adsorbates coupled by magnetic exchange ~\cite{FernandezRossier_PRL2009}, following an experimental study on the same system~\cite{HeinrichScience2006}. It considered the following Hamiltonian to describe the structure of the chain:
\begin{eqnarray} 
\hat{H}&=& \sum_i D \hat{S}_{i,z}^2  \nonumber \\
&+& E
( \hat{S}_{i,x}^2 - \hat{S}_{i,y}^2 ) +g_i \mu_B \vec{S_i} \cdot \vec{B} 
+ \sum_{i,j} J_{i,j} \vec{S_i} \cdot \vec{S_j}
\label{Rossier} 
\end{eqnarray} 
The sums over $i$ and $j$ run over the Mn sites in the chain. It thus
considers a set of local spins, $\vec{S_i}$, each interacting with
the surroundings via an anisotropy Hamiltonian and coupled together
by Heisenberg exchange couplings (in the present Mn case the coupling
between first neighbours is anti-ferromagnetic). Using the perturbation
approach to justify the $S^2$ formula (see section~\ref{S2}),
the conductance for a tip positioned above a given atom in the
Mn chain is obtained. These results are shown in Fig.~\ref{fig2a_prl_rossier_2010} for a
number of Mn atoms between 1 and 4 (taken from \cite{FernandezRossier_PRL2009}). As the main feature, this study confirms very well the
odd-even alternance of the conductance that has been observed
experimentally~\cite{HeinrichScience2006}. It appears as a direct
consequence of the spin structure of the chain: in a finite chain,
the ground state corresponds to a total spin equal to zero (5/2)
for the even (odd) numbers of atoms, leading to a very different
energy spectrum of excited states.  Another interesting feature
appears on the results for a chain of three atoms: the conductance
is different for the tip positioned above the central atom and above
the end atom; it concerns both the number of steps in the conductance
and their heights. This is linked with the symmetry of the problem:
a central excitation is symmetric, whereas exciting at the end of the
chain can be symmetric or antisymmetric. However, in the experimental
paper~\cite{HeinrichScience2006}, it is mentioned that the conductance does
not vary along the chain. The origin of this discrepancy is not clear.

The strong coupling approach has also been applied to the
problem of excitation of chains of magnetic atoms with an STM
tip~\cite{GauyacqPRB2011}. A model system was considered,
made of a spin-1/2 chain of atoms, with first neighbours
coupled by an Heisenberg exchange interaction (ferromagnetic
or anti-ferromagnetic). The formulation closely follows the one
outlined above in section~\ref{Strong}, except that the anisotropy
Hamiltonian describing the interaction between the local spin and its
environment is replaced by a Heisenberg chain interaction ($\sum_{i}
J \vec{S_i} \cdot \vec{S_{i+1}}$) representing the interaction of each local
spin with its neighbours. It is assumed  that tunnelling is very fast
so that the Heisenberg interaction can be neglected during tunnelling
and a sudden approximation can be used; it is also assumed that this
fast tunnelling only concerns one atom at a time, thus leading to the
definition of the total spin $S_T$ channels ($S_T$ = $S$ $\pm$ 1/2,
where $S$ is the local spin of one site of the chain). This model then
describes a chain of individual adsorbates,  only coupled via magnetic exchange.

The conductance of this system has been studied for various lengths of
the chains, with up to 18 atoms. The aim was to look for general trends
in the chain excitation, as well as to the conductance  behaviour when
the number of atoms in the chain goes to infinity, i.e. when spin wave
modes can be defined. Actually, it appears that finite chains with
a ring boundary condition (a Heisenberg chain of $N$ atoms with a
periodic boundary condition: $S_{N+1}$ = $S_1$) can be considered as
approximations of the infinite system thus allowing the discussion of
the characteristics of spin wave excitation by tunnelling electrons in
1D-Heisenberg chains. In other words, the spin wave modes confined in a
finite-size object, a ring, become quantized and they constitute
 a sub-set of the excited states in a finite size chain (note that
the number of states in a chain of $N$ atoms is equal to $2^N$ and
becomes extremely large when $N$ is large). 

In the ferromagnetic case, a finite-size ring of atoms can be considered
as a piece of the infinite system. In the ground state, all local spins
are aligned to form a state of maximum spin ($S_{Tot}$ = $N/2$, in the
present case with spin-1/2 sites). Excitation by a tunnelling electron
only populates the states with $S_{Tot}$ = $N/2$ -1, which  correspond to
the spin wave mode or 'magnon'. So, only  very specific states  among all the available excited magnetic states in the chain are actually excited by a tunnelling electron. The spin wave excitation is very efficient, similarly
to the case of a single adsorbate on the surface. 
From the analysis of the excited states of a ring that are populated by
tunneling electrons, it is possible to extract the $k$-distribution of
the excited waves ($k$ is the spin wave momentum). 
The $k$-spectrum of
the excited spin waves is white (all $k$-values are populated with equal
probability) which allows to derive simply the energy-spectrum of the
excited spin-waves and thus to fully characterize the spin-wave excitation
by a tunnelling electron~\cite{GauyacqPRB2011}. The equi-probability
of all $k$-states is a direct consequence of the nature of the
excitation process: the flip of a local spin in an infinite
ferromagnetic chain,
i.e. a $\delta$-like excitation.

\begin{figure}
\centering
\includegraphics[width=0.6\textwidth]{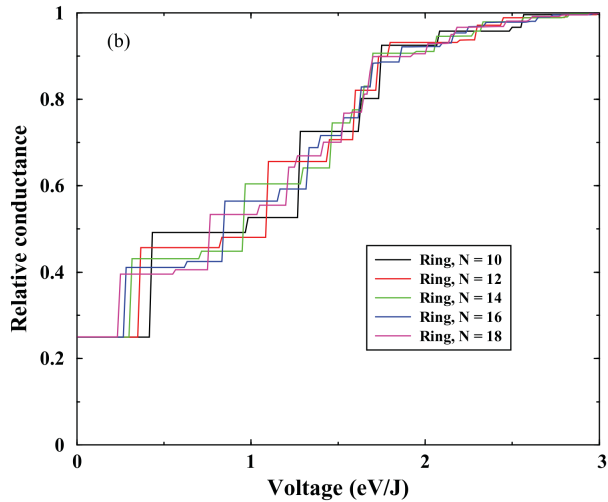}
\caption{\label{rel_cond_chain_anti}
Relative conductance of a ring of antiferromagnetic atoms as a function of the tip bias (in units of $J$). The conductance has been normalized to one above the inelastic magnetic thresholds. Five different numbers of magnetic atoms are presented: $N=10$ (black curve), $N=12$	(red curve), $N=14$ (green curve), $N=16$ (blue curve), and $N=18$ (magenta curve). \textcolor{red}{Taken from reference~\cite{GauyacqPRB2011}.}
}
\end{figure}

The anti-ferromagnetic case is quite different. First, a finite
size ring of atoms is only an approximation of the infinite system
due to strong correlations between distant spins (see discussions
in e.g.~\cite{DesCloizeaux,Orbach,Mattheis,GauyacqPRB2011}
and in text books~\cite{Yosida, Ashcroft}), so that one has to look
for the convergence of the conductance when $N$ goes to infinity. The
ground state of a chain with an even number of atoms corresponds to a
total spin $S_{Tot}$ equal to zero and the spin wave mode 
corresponds to  a set of $S_{Tot}$ = 1 states. Figure~\ref{rel_cond_chain_anti} shows
the conductance for a series of rings of different lengths, $N$ = 10 -
18  . It appears that a very large number of magnetic states in the ring
are excited by a tunnelling electron (there is a very large number of
steps in the conductance function of bias). All these excited state
contributions build a complex step function; however, if the number
$N$ of atoms in the ring is changed, the step positions and heights
change, but the continuous curve that would be obtained by smoothing
out the steps does not. Consequently,   the conductance functions with
many steps can be considered as approximations by step functions of a
continuous curve that  corresponds to the conductance of an infinite
chain, i.e. to the excitation of spin waves. It also appears that many
more states are excited, beyond the 'usual' spin wave mode,
in particular many states excited by tunnelling electrons belong to the
continuum of two-spinon states~\cite{KarbachPRB1997,KarbachPRB2000}
(two-spinons form a continuum and quantization in a finite size ring
transform them into discrete states forming several 'spin-wave modes'
at energies higher that the 'usual' low lying magnon mode).

\begin{figure}
\centering
\includegraphics[width=0.6\textwidth]{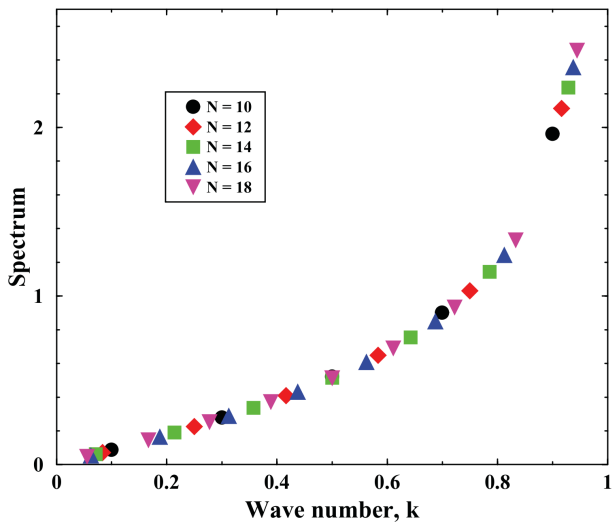}
\caption{\label{sum_spec_spin_waves}
Spectrum summed in energy of the spin waves
excited by a tunneling electron in a ring of antiferromagnetic atoms
as a function of their wave number $k$. The wave number is expressed
in units of $\pi /a$, where $a$ is the lattice spacing. Results for different
numbers of atoms in the ring are presented: see inset. \textcolor{red}{Taken from reference~\cite{GauyacqPRB2011}.}
}
\end{figure}

Figure~\ref{sum_spec_spin_waves}
 presents the $k$-distribution for various numbers
of atoms in the ring (see insert); it corresponds to the sum of all
the excited spin wave modes. Due to the finite size of the ring,
only a finite number of $k$-values are determined for each number N
of atoms. It appears that the convergence of the $k$-distribution with
the number of atoms in the ring is excellent, so that the distribution
in  Fig.~\ref{sum_spec_spin_waves} can be considered as the result for spin wave excitation in
an infinite 1D-chain.  It also appears that the distribution peaks at
the outer edge of the Brillouin zone, in contrast to the ferromagnetic
case where all $k$-values are equiprobable. This is connected with the
different qualitative views one can have of the excitation process in
the ferro and anti-ferromagnetic cases. In the  ground state of the
ferromagnetic chain,  all spins are aligned; only an incident electron
with an anti-parallel spin can excite the chain and the only excitation
it can induce is a spin-flip of the atom through which  tunnelling
occurs. As a consequence, in a ferromagnetic chain, only the spin wave
mode is excited with all $k$-waves equally populated. The situation is
different for anti-ferromagnets: due to the multiconfigurational
character of the spin wavefunction, a given atom in
the chain does not have a well defined spin. However, tunnelling of an
electron with a given spin through an atom selects a well defined $S_T$
spin coupling between the electron and the active site in the chain. For
example, if tunnelling implies the $S_T$ = 0 channel for a spin-1/2 atom,
this will select in the multiconfigurational expansion of the chain ground state
only the components that correspond to a spin of the adsorbate  opposite
to that of the tunnelling electron. 
\textcolor{red}{The tunnelling process thus selects a part of the ground state wavefunction; basically half of the ground state wavefunction is projected out  and this projection directly induces the excitation of many chain states. This correlation-mediated excitation is very broad and concerns many states. It emphasizes the correlations present in the chain ground state; in particular the two-atom period present in the chain correlation appears in the correlation-mediated excitation process and leads to a $k$-distribution peaked at the
edge of the Brillouin zone.}


This study of the excitation of Heisenberg chains has been extended to the case of frustrated ferro-magnetism in spin-1/2 chains~\cite{frustation2011}. Indeed, it has been shown that a Heisenberg chain with ferromagnetic coupling between first neighbours and anti-ferromagnetic coupling between second neighbours leads to a variety of magnetic structures when the relative magnitude of the exchange couplings is changed and when a magnetic field is applied; as said in Ref.~\cite{Hikihara2008}, 'the ground state phase
diagram ... is a zoo of exotic quantum phases' (see discussions in~\cite{Sudan2009,Kecke2007,Vekua2007,Chubukov1991} and references therein). Besides the fascinating properties of these structures, the theoretical activity in the field was partly prompted by experimental studies of various systems that were interpreted as frustrated 1D-spin 1/2 chains (see e.g.~\cite{Hase2004,Drechsler2007,Banks2009,  Tsirlin2011} and references therein). Such chains supported on a surface can be excited by tunnelling electrons and Ref.~\cite{frustation2011} studied how the variety of magnetic structures in the chain is reflected in the characteristics of the chain excitation by tunnelling electrons. Surprisingly, the changes in junction conductance following  changes in the chain magnetic structure was  found in most cases  not to be very strong. However, it was shown that the use of polarised electrons for exciting the chain could be used to identify structural features in the chains. In addition, in such frustrated chains, helical features of the ground state, remnants of the classical spiral structure, can only be seen in correlation functions; but  excitation of the chain by tunnelling electrons can reveal the underlying spiral stucture. This is connected with the excitation process at play in this system which bears much ressemblance with the process active in pure anti-ferromagnetic chains that was outlined above. Selection of a $S_{T}$ channel for the tunnelling process, selects only a part of the correlation expansion describing the chain ground state and this correlation-mediated process leads to a broad excitation spectrum; it also uncovers some hidden properties of the correlations in the chain, such as the helical feature, that become visible in the momentum spectrum of the spin waves excited by a tunnelling electron~\cite{frustation2011}.












\section{Kondo effect and IETS}
\label{Kondo}

The above sections have reviewed the very fast developments in magnetic
IETS with the STM since the first experiments~\cite{HeinrichScience2004}
 by Heinrich and co-workers in 2004. As a spectroscopy, the technique has
revealed the magnetic structure associated with single magnetic impurities
on a non-magnetic substrate. This is \textcolor{red}{deeply linked to } 
Kondo physics.
Indeed, Jun Kondo~\cite{JungKondo} showed that magnetic impurities in a
non-magnetic host undergo spin-flip scattering by the host's conduction
electrons leading to an increase of resistivity as the temperature
is lowered below a given value. The main difference with the magnetic
IETS mechanism expounded above is that the spin-flip scattering takes
place at zero energy: there is no exchange of energy  between
the incoming and outgoing electron, contrary to the above inelastic processes.

In the present section, we are going to restrict ourselves to the
link between the Kondo effect and IETS, without \textcolor{red}{dwelling upon} 
the very extensive literature associated to the Kondo
effect on surface and particularly its study with the STM~\cite{CrommieScience,Berndt,ternes}.

As temperature is lowered, the spin-flip scattering becomes coherent~\cite{Hewson}.
Electrons become correlated and at low temperatures the ground state
of the full system reveals the spin-flip process by screening out
the magnetic moment of the magnetic impurity. The ground state is a singlet.
As temperature is raised, electron-hole excitations become
available, and the system is not in its ground state any more. The
apperance of excitations lead to decoherence of the spin scattering
process and the impurity recovers its magnetic moment.

\begin{figure}
\centering
\includegraphics[width=0.6\textwidth]{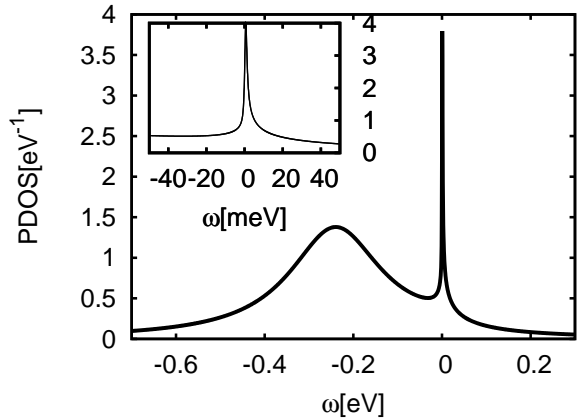}
\caption{\label{Kondofig}
Projected density of states on the magnetic adsorbate
electronic structure or spectral function $A$ as a function
of the electron energy $\omega$ with respect to the Fermi energy
of the substrate. 
The impurity has a single electronic level that gives rise
to the broad Lorentzian curve below the Fermi energy and the
electron-electron repulsion is very large so that the doubly
occupied state is far above the Fermi energy. Near the Fermi
energy the sharp Kondo resonance develops. \textcolor{red}{The insert presents an close view at the Fermi energy range. Taken from reference~\cite{KorytarCuPc}.}
}
\end{figure}

The electronic spectrum of such a Kondo system in its ground state is
characteristic of strongly correlated systems. We just saw that
the ground state is a singlet, and the first excitations recover
the local magnetic moment of the impurity. If we take the full electronic
spectrum of the system and keep the electronic structure that has
information on the impurity's state, we study the projected density
of states onto the impurity's orbitals also known as spectral function.
Figure~\ref{Kondofig} shows a typical Kondo spectral function
for the case of very large intra-impurity Coulomb repulsion.
At low temperature, this spectral function shows a broad peak that is largely
occupied and represents the Kondo ground state~\cite{Hewson,Fulde}. At
just the Fermi energy, a new peak appears that represents the first
excitations of the system. The first peak under the  Fermi energy
is the usual resonance of a localized electronic state in front of an
electronic continuun, it is sometimes called the charge fluctuation peak
to emphasize its origin in the electron hybridization between impurity and
substrate~\cite{Hewson,Fulde}. The second peak of low-energy excitations,
hence just above the Fermi energy, is the spin-fluctuation peak. This
peak is sometimes called the Abrikosov-Suhl resonance or simply, the
Kondo peak. The Kondo peak is a hallmark of the Kondo effect. Since it
is a spectral feature of the magnetic impurity, it can be easily revealed
by conductance measurements with the STM. Indeed, the STM conductance
is closely related to the local density of states. When the STM tip
is located at the impurity, the impurity's spectral function usually
dominates the conductance, and a peak at zero bias reveals the Kondo
effect.

Actual STM measurements are quite involved and the complete characterization
of the Kondo peak requires several measurements. However, the Kondo peak
has some unique features that makes it totally different from any other
source of a zero-bias anomaly. These feaures are:
\begin{enumerate}
\item The temperature dependence of the width of the peak. At very low
temperature the peak's width saturates to $2 k_B T_K$ where $T_K$
is the \textcolor{red}{so called } Kondo temperature and sets the energy scale at which
Kondo physics takes place. Hence, the peak's width, $\gamma$,  follows
the law~\cite{Nagaoka}
\begin{equation}
\gamma = 2 k_B \sqrt{(2 T_K)^2+ (\pi T)^2}
\label{width}
\end{equation}

\item The peak dependence with an external magnetic field, $\vec{B}$. 
An external magnetic field induces a Zeeman splitting of the impurity's
magnetic levels, hence spin up scattering is not any longer degenerate
with spin down scattering and the Kondo effect is destroyed. However,
its destruction is not complete, because electrons of higher energy
can interfer with opposite spin electrons if their energy difference
is exactly the Zeeman splitting. Indeed, we are describing an
inelastic effect and at the energy threshold of this inelastic
effect the electrons can spin-flip coherently recovering a
Kondo peak, but instead of being centered near the Fermi energy, the new peak
is centered at the energy of the excitation. Since electrons above
the Fermi energy and below (holes) can produce this inelastic effect,
there are two new peaks that appear replacing the previous Kondo
peak when the magnetic field is switched on. For very large
fields coherence cannot be achieved and all peaks disappear, or in
other words, the impurity's magnetic moment cannot be
screened out at large external magnetic fields.
 \end{enumerate}

\textcolor{red}{Both } the temperature and magnetic field behaviors permit us to
characterize a zero bias anomaly as due to the Kondo effect. However,
experiments are further complicated by the fact that these anomalies
are rarely pure peaks. Indeed, they present Fano lineshapes that
are more difficult to analyze. The origin of the Fano lineshape comes
from the contribution of many electronic states to the STM conductance 
other than the impurity's ones~\cite{Crommie}.

A further difficulty is the typical energy scale in the Kondo problem.
Rarely a Kondo temperature goes above 100K. Larger temperatures
typically mean that the charge fluctuation peak and the spin-fluctuation
peak start overlapping and a pure Kondo effect is no longer available:
in fact, charge fluctuations rapidly override spin ones, and the
system is a mixed-valence one, that fluctuates between several
charge states. The spectral features are very broad at the Fermi energy.

Due to the particular electronic structure of atomic and molecular
substrates the Kondo temperatures are anywhere between a few Kelvin and $\sim 100$ K.
This is in stark contrast with the initial quantum dot studies of
Kondo effects where typical temperatures where in the milli-Kelvin range
making it more difficult to measure. Usual low-temperature STM's are
capable of measuring adsorbate Kondo peaks. But they need to be able to change
their temperatures in a very controlled way within a few K to be able to
probe the above temperature dependence, Eq.~(\ref{width}).
They also need to be fitted with a magnetic field. However, due to the small
Bohr magneton value (57.9 $\mu$eV/Tesla) very large magnetic fields
are mandatory to be able to discern the above splitting of the Kondo peak.
At large Kondo temperatures, the splitting can be simply undetectable.

Besides these difficulties, the STM has become a successful \textcolor{red}{tool to reveal } 
coherent spin-flips in magnetic adsorbates.

\subsection{Inelastic spin-flip and Kondo effect}

As mentioned above, the impurity-substrate system in an excited
state does not exhibit Kondo physics anymore. However the transition
from a correlated ground state to an uncorrelated excited state is gradual.
Zar\'and and co-workers~\cite{Zarand,Borda} have undertaken the study of
this transition. 
The appearance of the Kondo ground state can be seen as the screening
of the magnetic impurity in such a way that the impurity behaves as
a strong but conventional potential scatterer where inelastic effects are
absent. Zar\'and and co-workers~\cite{Zarand,Borda} have studied the
energy dependence of the inelastic scattering
rate as the impinging electron energy is increased.
These authors take the total cross section of the electron-impurity scattering
and then substract the elastic contribution to the cross section, defining
in this way the inelastic part. By virtue of the optical theorem, the total
cross section is obtained from the imaginary part of the diagonal
T-matrix, while the elastic part is obtained from the on-shell
Golden-rule like expression using all terms from the T-matrix. They
obtained the T-matrix from a numerical renormalization group calculation
and use Fermi liquid identities to evaluate the absolute values of the 
electron cross seection.

They show that most of the scattering for electron energies above the
Kondo temperature, $T_K$, is inelastic. At energies below $T_K$ the
total cross section saturates, as well as the elastic one. The inelastic
cross section presents a maximum at about $T_K$. As the energy decreases
towards the Fermi energy (here taken as zero), the inelastic cross
section diminsishes with the square of the electron energy. This is
expected from Fermi liquid theory, since the quasiparticle
lifetime scales with the square of the electron energy close to the Fermi
energy. When the electron energy increases past $T_K$, the inelastic
cross section diminishes again, this time with $\sim 1/ln^2(\omega/T_K)$,
with $\omega$ the electron energy. This behavior is due to the dominance
of spin-flip scattering at large electron energies.

This study show how inelasticities destroy the Kondo effect, and permits
to characterize the onset of inelastic spin-flip scattering as
the electron energy increases.

\subsection{Magnetic IETS and Kondo effect}

Different  from the above case concerning the
 destruction of the Kondo ground state by inelastic processes is
the study of the impurity's magnetic structure with impinging electrons.
Zitko and Pruschke~\cite{Zitko:NJP2010} have
applied Kondo theories and described the coexistence of a Kondo peak
as well as IETS steps in the STM conductance of Co atoms on CuN/Cu(100)
substrates. Indeed, Kondo theories include information on incoherent
spin-flip scattering leading to IETS. Moreover, one can picture the IETS
process as a Kondo one at the excitation threshold because, at threshold,
the elastic and inelastic spin-flips are degenerate, and hence coherent.
It is then possible to obtain the same type of singular behavior as the
Kondo peak, but at the IETS thresholds. 

\begin{figure}
\centering
\includegraphics[width=0.4\textwidth]{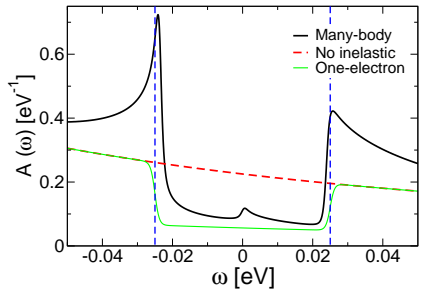}
\caption{\label{CuPc_singlet}
Projected density of states on the magnetic adsorbate
electronic structure or spectral function $A$ as a function
of the electron energy $\omega$ with respect to the Fermi energy
of the substrate. The vertical dashed lines show the one-electron
inelastic thresholds. \textcolor{red}{ Taken from reference~\cite{KorytarIETS}.}}
\end{figure}

This has been the subject of study by Hurley and collaborators~\cite{Hurley:ArXiv2011} and Koryt\'ar et al~\cite{KorytarIETS}. Hurley and collaborators~\cite{Hurley:ArXiv2011} use perturbation theory to analyze the magnetic IETS of Co and Fe on CuN/Cu(100).
The authors conclude that certain spike-like thresholds at the
steps of the experimental IETS are due to Kondo-like peaks. Koryt\'ar
et al~\cite{KorytarIETS} use a self-consistent approach to study the
excitation of a singlet molecule to its triplet state. This type
of singlet-triplet excitations have already been studied
experimentally in Mn chains with an even number of atoms~\cite{HeinrichScience2006}, in carbon nanotubes~\cite{PaaskeNatPhys2006}, in C$_{60}$ molecules~\cite{RochNature2008},
and in cobalt complexes~\cite{ParksScience2010}.

Koryt\'ar and collaborators~\cite{KorytarIETS} explicitly include an exchange
interaction between two localized electrons that belong to different
orbitals. This model has been used to explain the
experimental behavior of copper phthalocyanine on Ag (100), showing
that the Kondo system is related to a triplet-singlet
spin transition~\cite{KorytarCuPc} as shown in the experiments by
Mugarza  et al~\cite{Mugarza}. The exchange interaction is then:
\begin{equation}
\hat{H}_I = I \vec{S}_1 \cdot \vec{S}_2.
\label{HI}
\end{equation}
For positive $I$, the impurity is a singlet and it is not subjected to
the Kondo effect. As a result there is no peak at zero energy. 
The triplet state lies at exactly $I$ above the singlet state in this
model. However,
as the electron energy increases, the triplet state becomes available
under an inelastic spin-flip process. At threshold, both spin states
become degenerate and spin-flips keeps its coherence giving rise
to singular-like peaks. Figure~\ref{CuPc_singlet} shows the
spectral function for such a system. In the same graph, the above
one-electron results for magnetic IETS is also plotted. We see
that many-body effects included via the self-consistent approach
substantially change the spectral function. The most noticeable
feature is the new peaks that considerably change the overall shape
of the spectral function. But also the excitation thresholds
are perturbed. The thresholds are shifted to lower energies. This
is an important effect for the determination of magnetic energies
based on IETS: many-body effects can renormalize the IETS thresholds.


The strength of the threshold renormalization is found to depend on
an energy scale that replaces the Kondo temperature and is called
$T_K^0$ in Ref.~\cite{KorytarIETS}. Indeed\textcolor{red}{, if $I=0$, a Kondo
effect appears} with Kondo temperature $T_K^0$. Another parameter
of the strength of the threshold renormalization is
$I$. 
It is interesting to study the change in the excitation energy,
called $\Delta I$, as a function of $I$. For large $I$, the threshold renormalization, $\Delta I$, follows a $I/ln(I/T_K^0)$ behavior,
 similar to the renormalization
found for the shift of Kondo peaks with an external magnetic field~\cite{MoorePRL2000,RoschPRB2003}. The resemblance of the present results with
the ones found in the presence of magnetic fields is due to
the similar physical process: in both cases, there is a magnetic
excitation, in the present case due to the interaction between
two localized spins, and in the magnetic-field case due to Zeeman
energy splitting, and when the electron energy is large
enough to open the excited channel, the ground and excited states
are connected via spin flip.

The emergence of the excitation energy, $I$, as a new energy scale
for Kondo physics in the presence of magnetic excitations, can be
further revealed by studying the impurity's spectral function behavior
with temperature~\cite{KorytarIETS}. When $I=0$, the system displays
a characteristic Kondo temperature, named $T_K^0$ above. At temperatures
$T$ larger than $T_K^0$, the Kondo peak is very diminished. However,
when $I \neq 0$, the new Kondo peaks appearing at the excitation
thresholds persists well beyond $T_K^0$, showing that a new type of Kondo physics
is appearing. When the temperature matches $I$, the Kondo peaks
at the excitation thresholds coalesce and there is a unique Kondo
peak that persists at higher temperatures. Hence $I$ becomes
an important energy scale when considering Kondo effects.

The physics behind this new energy scale can be understood by looking
at the effect of thermal electron-hole pairs. For $I=0$, both
initial and final electrons have the same energy after scattering off
the magnetic impurity. Hence, both states are subject to the
decoherent effect of thermal electron-hole pairs near the Fermi energy.
As a consequence, beyond $T_K^0$, decoherence becomes very
large and the Kondo peak vanishes. When $I \neq 0$, the initial
electron is at $I$ away from the Fermi energy, if the
temperature is smaller than $I$, then these electrons will not be
affected by the decoherence induced by thermal electron-hole pairs. As
a result, the new Kondo physics is more resistant to higher temperatures.

\section{Conclusions}

The extraordinary \textcolor{red}{extension} 
of the STM to very low temperatures \textcolor{red}{with} built-in
magnetic fields has permitted the development of magnetic inelastic
electron tunneling spectroscopy with subatomic resolution. New
measurements have revealed the low-energy scale associated with
the magnetic anisotropy of adsorbates on solid surfaces. Atomic
adsorbates have been explored on different metallic surfaces either
with an insulating layer to decouple the adsorbate, such as MgO~\cite{HeinrichScience2004} or CuN~\cite{HeinrichScience2006,HeinrichScience2007}, or
on the surface itself~\cite{BalashovPRL2009,KhajetooriansPRL2011}. 
Also, IETS \textcolor{red}{proved to be } 
extremely useful in the study and
characterisation of magnetic impurities on semiconducting substrates~\cite{K}. Non-atomic adorbates such as magnetic molecules~\cite{TsukaharaPRL2009,ChenPRL2008} and
layers of magnetic molecules have also been studied revealing not only
the intramolecular magnetic properties but also intermolecular interactions.
All these experiments show that magnetic IETS is a full-fledged, versatile
technique of extreme usefulness. Its older sibling, vibrational IETS~\cite{StipeScience1998,Ho2002} has been credited with enhancing the STM atomic
resolution to chemical resolution, and now, magnetic IETS \textcolor{red}{brings STM to magnetic resolution at the atomic scale.}

Spurred by the experimental success and evolution, several \textcolor{red}{new theoretical approaches emerged.}
Initially, the first theories were of perturbational
character~\cite{Fransson2009,FernandezRossier_PRL2009}
 and explained the experimental scaling with the
spin operator~\cite{HeinrichScience2007}.
Beyond perturbation theories, two sudden-approximation approaches
have been very succesful. The first one deals with the complete
atomic T-matrix~\cite{Mats_PRL2009} and explains the spin-operator
dependence of the experimental data. The second one, uses the
very large spin coupling to 
\textcolor{red}{enforce the spin symmetries (value of the total spin, sum of the tunnelling electron spin and of the adsorbate spin) in the tunnelling process } 
~\cite{Lorente_PRL2009}. In this way, the excitation is seen
as the \textcolor{red}{sharing} 
of the incident electron flux among all possible
final channels, where \textcolor{red}{the} initial and final channels are connected
via the formation of \textcolor{red}{states of a well-defined total spin symmetry. 
In this way, the main features of the magnetic excitation process can be derived simply from spin coupling coefficients~\cite{Lorente_PRL2009}}

All of the above approaches are one-electron. 
These one-electron treatments are easy to implement using
first-principles input and they generally yield
quantitative data.
However,
many-body approaches are necessary when dealing with spin-flip dynamics. Indeed,
Kondo-related physics has to be considered and this can have
some important consequences on IETS such as the renormalization of
the inelastic thresholds~\cite{KorytarIETS}.

We foresee a lot of new developments in this field when
atomic manipulation techniques \textcolor{red}{are} 
combined with magnetic
IETS. On one side, new structures, taylored to \textcolor{red}{exhibit special features~\cite{OttePRL2009,KhajetooriansScience2011} 
 in view of peculiar 
applications will be searched for.} 
On
the other side, dynamical studies of magnetic \textcolor{red}{structures (e.g. direct study of the time evolution of a local magnetic moment) are already possible~\cite{LothScience2010}.} 
The combination of several techniques with 
magnetic IETS will surely give rise to new data and exciting applications.
We think about noise studies combined with IETS~\cite{Johannes}
as well as resonance study with $\mu$-wave probes~\cite{KomedaAPL2008,Balatsky}. The
ongoing development of the dynamical aspects of STM, together
with all the know-how achieved these \textcolor{red}{past years will} 
\textcolor{red}{further bring new exciting achievements } 
in the field.
These developments will \textcolor{red}{prompt} 
theoreticians to develop
non-equilibrium techniques to treat the excitations \textcolor{red}{with many body interactions taken into account, in parallel to} 
 the ever growing need for more quantitative, first-principles
based calculations. 

\bibliography{references_Lorente} 
\end{document}